\documentclass[twocolumn,showpacs,amsfonts,aps,prc,floatfix,%
nofootinbib,float]{revtex4} 
\usepackage{amsmath}
\usepackage{bm}
\usepackage{graphicx}

\newcommand{\be}[1]{\begin{equation}\label{#1}}
\newcommand{\ee}{\end{equation}}

\usepackage{color}

\begin{document}
%%%%%%%%%%%%%%%%%%%%%%%%%%Front Matter%%%%%%%%%%%%%%%%%%%%%%%%%%%%%%%%%%
%%%%%%%%%%%%%%%%%%%%%%%%%%%%%%%%%%%%%%%%%%%%%%%%%%%%%%%%%%%%%%%%%%%%%%

\title{Heavy Quark Diffusion with Relativistic
 Langevin Dynamics \\ 
 in the Quark-Gluon Fluid}
\date{\today}

\author{Yukinao Akamatsu}
\affiliation{Department of Physics, The University of Tokyo, Tokyo 113-0033, Japan}

\author{Tetsuo Hatsuda}
\affiliation{Department of Physics, The University of Tokyo, Tokyo 113-0033, Japan}

\author{Tetsufumi Hirano}
\affiliation{Department of Physics, The University of Tokyo, Tokyo 113-0033, Japan}

\begin{abstract}
 The relativistic diffusion process of heavy quarks  
 is formulated on the basis of
 the relativistic Langevin equation in It\^{o} discretization scheme.
   The drag force inside the
  quark-gluon plasma (QGP) is parametrized according to the formula for the strongly
   coupled plasma obtained by
   the  AdS/CFT correspondence.  The diffusion dynamics of
  charm and bottom quarks in QGP is described by 
   combining the Langevin simulation under the background matter 
  described by the  relativistic hydrodynamics.
  Theoretical calculations of 
   the nuclear modification factor $R_{\rm{AA}}$ and the elliptic flow
   $v_{2}$ for  the single electrons from the charm and bottom decays
   are compared with the experimental data from the relativistic heavy ion collisions.
  The $R_{\rm{AA}}$ for electrons with large transverse momentum
     ($p_{\rm T} > 3$ GeV) indicates that the drag force from the QGP
      is as strong as the AdS/CFT prediction.

\end{abstract}


\maketitle

%%%%%%%%%%%%%%%%%%%%%%%%%%%%%%%%%%%%%%%%%%%%%%%%%%%%%%%%%%%%%%%%%%%%%%%
\section{Introduction}
\label{sec1}
\vspace*{-2mm}
%%%%%%%%%%%%%%%%%%%%%%%%%%%%%%%%%%%%%%%%%%%%%%%%%%%%%%%%%%%%%%%%%%%%%%%
 The physics of the 
 quark-gluon plasma (QGP) 
 is actively studied 
  by means of the relativistic heavy-ion collisions at Relativistic Heavy Ion Collider
   (RHIC) in BNL and will be pursued further at 
   Large Hadron Collider (LHC) in CERN  \cite{QGP}.
 The space-time evolution of the heavy-ion collisions at RHIC is well described
 by the (3+1)-dimensional relativistic  hydrodynamics
 supplemented with the hadronic cascade after chemical freezeout \cite{Hirano:2008hy}.
 Information on the collective dynamics of QGP is obtained 
 by the soft probes such as distributions of light hadrons at low momentum, while
  the information of 
 microscopic dynamics of QGP is obtained by
 the hard probes  such as jets, heavy quarks, and heavy quarkoniums \cite{Vitev2008}.

 In the present paper,
 we focus on charm and bottom quarks which behave as impurities
in QGP. Experimentally, the signal of the heavy quarks can be extracted from the single 
 electron spectra through semileptonic decays \cite{phenix2007, star2007}.
 Theoretically,  the energy loss of heavy quarks in QGP 
 has been estimated in  perturbative QCD (pQCD) techniques
  \cite{Djordjevic2006,Wicks2007}.
 However, it was pointed out recently that
  the convergence of the weak coupling expansion of the drag force for  
   heavy quarks is rather poor at the temperature relevant to RHIC and LHC, so
  that the calculation in the leading order  would not be
  reliable  \cite{Moore2008}.  Possible alternative way to estimate the 
   drag force is to use the duality conjecture between the 
   gauge theory and string theory (AdS/CFT correspondence) \cite{Herzog2006,Gubser2006,Teany2006}. Although it 
   can be applied only to the 
  ${\cal N}=4$ supersymmetric Yang-Mills plasma
  with large  't Hooft coupling,  
  the result obtained may provide us with 
  a hint for the drag force in the strong coupling regime of the  QCD plasma
  if appropriate translation is made \cite{Gubser2007}.

 The purpose of this paper is to study the connection between the drag force
 acting on the charm and bottom quarks in QGP  
 and  the final electron spectra.
 To make such connection, we introduce relativistic
  Langevin equation for heavy quarks under the background of the quark-gluon fluid described
   by the ideal hydrodynamics.
   We need relativity since the transverse momentum of the 
    heavy quarks at RHIC is not
     necessarily smaller than their rest masses.
  Our relativistic Langevin equation is formulated in It\^{o}  discretization scheme.
 The diffusion constant and the drag force are related through
   a generalized fluctuation-dissipation relation. As for the drag force,
    two distinct models, pQCD and AdS/CFT, are considered.  
    To calculate the space-time dynamics of light quarks and gluons, (3+1)-dimensional
    hydrodynamics is used, which is necessary to calculate the electron spectra 
    of different impact parameters in the heavy ion collisions.
  The Langevin  equation for heavy quarks is numerically solved from the initial
   distribution generated by Monte Carlo generator  PYTHIA \cite{Sjostrand} 
      until the freezeout of the heavy quarks. The transverse-momentum 
      ($p_{\rm T}$) dependence of the
      nuclear modification factor ($R_{\rm AA}$) and the elliptic flow ($v_2$) 
      of single electrons as decay products of 
     heavy quarks are calculated and compared with the RHIC data. 
 
 This paper is organized as follows.
In Sec.~\ref{sec2}, after formulating the 
  relativistic Langevin equation and a generalized fluctuation-dissipation
   relation, we introduce two extreme models of the drag force motivated by
   pQCD and AdS/CFT.
 In Sec.~\ref{sec3}, the relativistic hydrodynamics for light degrees of freedom
  and the relativistic Langevin equation for heavy degrees of freedom are
   combined in order to describe the heavy quark diffusion in dynamical QGP fluid.
 The initial condition of heavy quarks, the algorithm of Langevin simulation in dynamical background, and the treatment of the  freezeout and decay of heavy quarks are discussed
  in detail.
In Sec.~\ref{sec4}, the numerical results of our calculation are presented.
We show the profile of heavy quark diffusion, heavy quark spectra, and single electron spectra and compare our single electron spectra with experimental data.
Physical meaning of our results  are also discussed.
Section V is devoted to summary and concluding remarks.
In Appendix A, we show a derivation of the relativistic Kramers equation from the
 relativistic Langevin equation in the It\^{o}  discretization scheme. 

%%%%%%%%%%%%%%%%%%%%%%%%%%%%%%%%%%%%%%%%%%%%%%%%%%%%%%%%%%%%%%%%%%%%%%%
\section{Langevin Dynamics of Heavy Quarks}
\label{sec2}
%%%%%%%%%%%%%%%%%%%%%%%%%%%%%%%%%%%%%%%%%%%%%%%%%%%%%%%%%%%%%%%%%%%%%%%
In this section, we formulate the relativistic Langevin equation
 in the local rest frame of the fluid. A generalized form of the 
 fluctuation-dissipation relation is derived. Then, we discuss the
  drag forces calculated in the  pQCD approach and
   in the AdS/CFT approach. Finally, we introduce a phenomenological model
   of the drag coefficient $\Gamma$ and the diffusion constant $D$ which
   satisfy the generalized fluctuation-dissipation relation.    

%%%%%%%%%%%%%%%%%%%%%%%%%%%%%%%%%%%%%%%%%%%%%%%%%%%%%%%%%%%%%%%%%%%%%%%
\subsection{Relativistic Brownian motion}
\label{sec2a}
%%%%%%%%%%%%%%%%%%%%%%%%%%%%%%%%%%%%%%%%%%%%%%%%%%%%%%%%%%%%%%%%%%%%%%%
 
 Suppose that there exist a time scale $\tau_{\rm B}$ during which a Brownian 
 particle changes  
   its momentum and  a microscopic time scale $\tau_{\rm m}$
  during which light dynamical degrees of freedom change state and lose time correlation. 
   If  these time scales satisfy $\tau_{\rm B}\gg\tau_{\rm m}$, one 
  can describe the Brownian particle by the Langevin equation \cite{Langevin}.
  In such a situation that the charm/bottom diffuses inside the quark-gluon plasma,
  there are some extra complications: 
  (i) The background quark-gluon fluid 
  expands rapidly in space and time with the local 4-velocity $u^{\mu}(\vec{x},t)$,
   and (ii) the initial momentum distribution
  of the charm/bottom governed by the hard QCD process has high momentum
  component larger than their quark masses.  
 
  As for (i), we define  the Langevin equation in the rest frame of matter 
  ($u^{\mu}=(1, 0,0,0)$) and the
   real motion of the Brownian particle is obtained by
    the local Lorentz boost back to the moving frame.  As for (ii),
  we take into account the relativistic kinematics of the Brownian particle
  in a minimal way by using relativistic dispersion relation $E(p)=\sqrt{p^2+M^2}$.  
  Then the  Langevin equation in the rest frame of matter with minimum relativistic
  kinematics may be written as  \cite{Debbasch}
\begin{eqnarray}
\label{eq:dx}
\Delta\vec x(t)&=&\frac{\vec p}{E(p)} \Delta t , \\
\label{eq:dp}
\Delta\vec p(t)&=&-\Gamma(p) \vec p \Delta t + \vec \xi (t) .
\end{eqnarray}
Here  $\Delta \vec{x} (t) = \vec{x}(t') - \vec{x}(t)$,
  and $\Delta \vec{p} (t) = \vec{p}(t') - \vec{p}(t)$,
   and $\Delta t \equiv t ' -t$ is a discrete step of time.
 The momentum dependent drag coefficient is denoted by
 $\Gamma(p) $ which is related to the time scale of the Brownian particle 
$\tau_{\rm B} \sim \Gamma ^{-1}$.
Also $\vec \xi(t)$ is a noise 
 obeying the probability distribution $W[\xi(t)]$ which
 we take to be the Gaussian white noise with a normalization constant $C$:
\begin{eqnarray}
W[\vec\xi(t)]d^{3}\xi(t)=C
\exp \left[ -\frac{\vec \xi(t)^2 }{2D(p) \Delta t }  \right]
d^{3}\xi(t) .
\end{eqnarray}
This leads to
\begin{eqnarray}
\langle \xi_i(t)  \rangle &=&0 , \\
 \langle \xi_{i}(t) \xi_{j}(t') \rangle  
 &=& D (p) \delta_{ij} \delta_{tt'} \Delta t ,
\end{eqnarray}
where $D(p)$ is a momentum dependent diffusion constant.
Note that $\vec \xi(t) $ is not a single microscopic kick but a 
sum of microscopic  kicks during the time $\Delta t$.  

Throughout this paper, we use the It\^{o} discretization scheme of the 
 Langevin equation;  namely, all the argument of $\vec{p}$ in the 
right hand side of  Eqs.~(\ref{eq:dx}) and (\ref{eq:dp}) are evaluated at the 
 pre-point time $t$. This is particularly useful for numerical simulations
  due to obvious reason.
 The relativistic Kramers equation, which is a partial differential equation for the 
probability of the particle distribution in the phase space 
$P(\vec p,\vec x,t)$, is then obtained as (see the derivation in Appendix A)
\begin{eqnarray}
& & \! \! \! \! \! \! \! \! 
\left( \frac{\partial}{\partial t}+\frac{\vec p}{E}\frac{\partial}
{\partial \vec x} \right) P(\vec p,\vec x,t) \nonumber \\
& & = \frac{\partial}{\partial \vec p}\Bigl(
\Gamma(p) \vec p + \frac{1}{2}\frac{\partial}{\partial \vec p}
D(p) \Bigr)P(\vec p,\vec x,t). 
\label{eq:Kramers}
\end{eqnarray}
Note that $\partial/\partial \vec p$ acts not only on $P$ but also on
 $\Gamma(p)$ and $D(p)$.

Demanding that Eq.~(\ref{eq:Kramers}) has the relativistic
 Maxwell-Boltzmann distribution (the J\"{u}ttner distribution)
 $P(\vec p,\vec x,t)\propto \exp[-\sqrt{p^2+M^2}/T]$ as 
 a stationary solution, we obtain a constraint between the drag and the 
 diffusion as
\begin{eqnarray}
\Gamma(p) + G(p) = \frac{D(p)}{2ET},
\label{eq:FDT}
\end{eqnarray}
with $G(p) \equiv dD(p)/2pdp=dD(p)/d(p^2)$. 
If  $D$ is $p$-independent,
 Eq.~(\ref{eq:FDT}) reduces to 
  the relativistic analogue of the Einstein relation 
  $ \Gamma = \frac{D}{2ET}= \frac{M}{E} \frac{D}{2MT}$ 
   obtained in \cite{Debbasch}.

%%%%%%%%%%%%%%%%%%%%%%%%%%%%%%%%%%%%%%%%%%%%%%%%%%%%%%%%%%%%%%%%%%%%%%%
\subsection{Modeling the energy loss of heavy quarks}
\label{sec2b}
%%%%%%%%%%%%%%%%%%%%%%%%%%%%%%%%%%%%%%%%%%%%%%%%%%%%%%%%%%%%%%%%%%%%%%%

 Energy loss of heavy quarks in the deconfined phase
 has two sources; 
  the collisional energy loss due to elastic scattering of a 
  heavy quark with the plasma constituents and
   the radiative energy loss associated with the induced emission of the gluon.
    In the leading order (LO) of the 
  weak-coupling QCD perturbation, these processes are 
   found to have  different momentum dependence of the heavy quark
   and could become comparable in magnitude \cite{Djordjevic2006,Wicks2007}.  
  Recently, the convergence of such weak-coupling expansion was questioned
  by an explicit calculation of the  
  collisional process in the next-to-leading order (NLO) \cite{Moore2008}: The 
  drag coefficient for the 3-color, 3-flavor QCD 
  in the non-relativistic kinematics ($M \gg T, p$) reads
 \begin{eqnarray}
& &\Gamma_{\rm pQCD} |_{M \gg T, p}  \nonumber \\
& & \ \ \ \simeq \frac{8 \pi}{3}\alpha_{\rm s}^2 \frac{T^2}{M}
  \left( - \ln g + 0.07428 + 1.8869 g \right)  .
 \label{eq:G-pQCD}
 \end{eqnarray}
 For the QCD coupling constant relevant at RHIC and LHC ($ g \sim 2$),
 the weak-coupling expansion has an obvious problem of convergence.
 From the phenomenological point of view, it has been argued in the past
 that relatively large  drag force is necessary to account for the
 RHIC data \cite{Moore:2004tg,vanHees:2004gq}.

  Alternative  approach to the drag force is provided by the 
  AdS/CFT correspondence \cite{Herzog2006,Gubser2006,Teany2006}.
  In the  $\mathcal N=$ 4 
 super-Yang-Mills theory (SYM) at large $N_{\rm c}$ and large 't Hooft coupling
  $\lambda \equiv g^{2}_{_{\rm SYM}}N_{\rm c}$,
 energy loss of an external quark with velocity $v$ is obtained as
 \begin{eqnarray}
 \frac{dp}{dt} &=&
 -\frac{\pi \sqrt{ \lambda} }{2}T_{\rm{SYM}}^{2}\frac{v}{\sqrt{1-v^2}}  
 \label{eq:SYM_drag} \\
 &\simeq &-\frac{\pi \sqrt{ \lambda} }{2}T_{\rm{SYM}}^{2}\frac{p}{M}. 
\end{eqnarray}
Here the first equation is valid for arbitrary mass of external quark,
 while the second equation is valid for $M \gg \sqrt{\lambda} T$ \cite{Herzog2006}.  
 By  matching the energy density and the heavy quark potential
  in  the SYM plasma to those in QCD plasma, 
 one finds
 $ T_{\rm{SYM}} \simeq {T_{\rm{QCD}} }/{3^{1/4}}$ and 
 $3.5 \leq  \lambda \leq  8.0$  \cite{Gubser2007}. Then, 
 the drag coefficients may be estimated as 
\begin{eqnarray}
\Gamma_{\rm AdS/CFT} = (2.1 \pm 0.5) \frac{T^2}{M}.
\label{eq:G-SYM}
\end{eqnarray}
 A remarkable feature of this formula in contrast to the
 weak-coupling estimate is that
 $\Gamma$ is $p$-independent \cite{Herzog2006}.
 The fundamental question here is, of course,  the reliability of the
  translation from SYM to QCD both conceptually and 
   numerically.

Given the theoretical uncertainties in estimating the 
drag force as mentioned above,
 we will take a phenomenological approach in this paper:
 We adopt the parametric dependence of the drag coefficient
 motivated by the AdS/CFT in Eq.~(\ref{eq:G-SYM}) with the overall 
 magnitude left as a free parameter:
\begin{eqnarray}
\Gamma \equiv \gamma \frac{T^2}{M}.
\label{eq:G-param}
\end{eqnarray}
 The dimensionless drag coefficient $\gamma$ is
 assumed to be independent of $T, M,$ and $p$ throughout this paper.
  The corresponding diffusion constant $D$ is obtained from the
  generalized fluctuation-dissipation relation
   in Eq.~(\ref{eq:FDT}) with the physical boundary condition,
    $D \rightarrow 0 $ as $\Gamma \rightarrow 0$: 
\begin{eqnarray}
D = 2 E T \cdot \Gamma \cdot  \left( 1 + \frac{T}{E} \right)
 =  \gamma \frac{2T^3}{M}(E+T) .
\label{eq:D-param}
\end{eqnarray}   

It is in order here to make two remarks on the
dynamics we employed in Eqs.~(\ref{eq:G-param}) and (\ref{eq:D-param}).
(i) Since we have assumed $\Gamma$ to be $p$-independent
 motivated by AdS/CFT,
 the diffusion constant $D$ depends necessarily on the momentum of
 the  Brownian particle.  One may alternatively assume
 that $D$ is independent of $p$ while $\Gamma$ depends on $p$
 as $\Gamma(p) = D/(2 E(p)T)$ \cite{Debbasch}.  Such dynamics 
 would simulate the $p$-dependence of the drag force
 due to collisional process in the weak-coupling regime 
 \cite{Wicks2007}.
 (ii) At ultra-high energies $p \gg M$, the dominant energy loss
  occurs through  the induced 
  emission of the gluons.  In this case,  
  the condition $\tau_{\rm B} \gg \tau_{\rm m}$ 
  is violated.
Thus the Langevin approach becomes
   inapplicable \cite{Herzog2006} and a 
   different approach based on radiative energy loss is required
  to describe heavy quarks in the QGP \cite{Wicks2007,Armesto:2005mz}.
  With these reservations in mind, we consider our ansatz 
  Eqs.~(\ref{eq:G-param}) and (\ref{eq:D-param}) 
  as phenomenological but characteristic dynamics of QCD and
  try to estimate  $\gamma$ from
   the observed single electron data at RHIC in later
   sections.

%%%%%%%%%%%%%%%%%%%%%%%%%%%%%%%%%%%%%%%%%%%%%%%%%%%%%%%%%%%%%%%%%%%%%%%
\section{Hydro + Heavy-Quark Model}
\label{sec3}
%%%%%%%%%%%%%%%%%%%%%%%%%%%%%%%%%%%%%%%%%%%%%%%%%%%%%%%%%%%%%%%%%%%%%%%

%%%%%%%%%%%%%%%%%%%%%%%%%%%%%%%%%%%%%%%%%%%%%%%%%%%%%%%%%%%%%%%%%%%%%%%
\subsection{Background QGP fluid}
\label{sec3a}
%%%%%%%%%%%%%%%%%%%%%%%%%%%%%%%%%%%%%%%%%%%%%%%%%%%%%%%%%%%%%%%%%%%%%%%

The hydrodynamics  has been quite successful
in the description of collective flow phenomena in heavy ion collisions
at RHIC. Since the hydrodynamics
  gives space-time evolution
of temperature and flow velocity of the fluid so that 
 the local rest frame of fluid is well-defined.
 Then,  the Langevin equation in 
 the previous section formulated in the local rest frame of the fluid
  is applicable directly. 

Let us first summarize the relativistic
hydrodynamic model \cite{Hirano:2008hy,Hirano:1,Hirano:2,Hirano:had_cas}
whose basic equation reads
\begin{eqnarray}
\label{eq:hydro}
\partial_{\mu}T^{\mu\nu}&=&0.
\end{eqnarray}
Here $T^{\mu\nu}$ is energy-momentum tensor.
For strongly interacting matter with zero viscosity,
$T^{\mu\nu}$ becomes
\begin{eqnarray}
\label{eq:idealdecom}
T^{\mu\nu}&=&(e+P)u^{\mu}u^{\nu}-Pg^{\mu\nu},
\end{eqnarray}
where $e$, $P$, and $u^{\mu}$ are energy density, pressure, and
four fluid velocity, respectively.
The baryon chemical potential is neglected, because it is small 
near mid-rapidity at RHIC energies.
We solve Eq.~(\ref{eq:hydro})
in the Bjorken coordinates ($\tau$, $x$, $y$, $\eta_s$),
where $\tau = \sqrt{t^2-z^2}$ and
$\eta_s = \frac{1}{2}\ln[(t+z)/(t-z)]$
are proper time and space-time rapidity, respectively. 

In the high temperature ($T>T_{c}=170$ MeV) QGP phase, we employ
the bag equation of state (EOS) for massless partons
($u$, $d$, $s$ quarks and gluons) with $B^{1/4}=247.19$ MeV.
Here the bag constant is tuned to have transition
to the hadron resonance gas at $T_{c}$.
In the hadron phase ($T<T_{c}=170$ MeV),
a resonance gas of hadrons with the mass up to $\Delta(1232)$
is employed \cite{Hirano:2}.
 Volume fraction of QGP $f_{\mathrm{QGP}}$ 
in the mixed phase is
\begin{eqnarray}
\label{eq:frac_qgp}
f_{\mathrm{QGP}} = \frac{e-e_{\mathrm{had}}}{e_{\mathrm{QGP}}-e_{\mathrm{had}}} ,
\end{eqnarray}
where $e_{\mathrm{QGP}}$ ($e_{\mathrm{had}}$)
is the maximum (minimum) value of the energy density in the mixed phase.
Later we will utilize $f_{\mathrm{QGP}}$ to define the effective lifetime of QGP
and the freezeout condition for the heavy quarks.

 Hot QGP with local thermalization is assumed to be 
 produced at $\tau_{0}$ = 0.6 fm. 
 The entropy density distribution at $\tau_0$ in the mid-rapidity
 is taken to be proportional to a linear combination of 
 the number densities of participants and binary collisions
 in the transverse plane \cite{Hirano:had_cas}.
For the
 initial condition of the flow velocity, Bjorken's scaling solution, 
 $u_{x}(\tau_{0}) = u_{y}(\tau_{0}) =0$ and
$u_{z}(\tau_{0}) = \sinh \eta_{s}$, is  employed \cite{Bjorken}.
With these initial conditions, the hydrodynamic model can well reproduce the experimental data of charged particles at RHIC
\cite{Hirano:2008hy}.

The space-time evolution of the QGP fluid obtained as above
has been exploited for a quantitative study of hard and rare
probes such as azimuthal jet anisotropy, nuclear modification factor
of identified hadrons, disappearance of back-to-back jet correlation,
$J/\psi$ suppression, and direct photon emission~\cite{HiranoCollectPapers}.

%%%%%%%%%%%%%%%%%%%%%%%%%%%%%%%%%%%%%%%%%%%%%%%%%%%%%%%%%%%%%%%%%%%%%%%
\subsection{Heavy quark diffusion in quark-gluon fluid}
\label{sec3b}
%%%%%%%%%%%%%%%%%%%%%%%%%%%%%%%%%%%%%%%%%%%%%%%%%%%%%%%%%%%%%%%%%%%%%%%
We solve the Langevin equation (\ref{eq:dx}) and (\ref{eq:dp})
with the drag coefficient $\Gamma$ given by Eq.~(\ref{eq:G-param}) 
in the local rest frame of the fluid element.
The dimensionless parameter $\gamma$ is inversely proportional to the relaxation 
time $\tau_{\mathrm{Q}}$ of a heavy quark as 
\begin{eqnarray}
\tau_{\mathrm{Q}} = \frac{1}{\Gamma}
= \frac{M_{Q}}{\gamma T^{2}}. 
\end{eqnarray}
The $\tau_{\mathrm{Q}}$ is listed in Table \ref{RELAX_LIFE}(a) for three
 typical values, $\gamma=0.3$ (weak coupling), $\gamma=1.0$ (intermediate coupling),
   and $\gamma=3.0$ (strong coupling).  The characteristic temperature
   felt by the heavy quark during the space-time history 
    in the QGP fluid is taken to be 210 MeV (as for the reasoning of this 
    number, see  Sec.~\ref{sec4a1}).

 Let us now introduce an effective lifetime of QGP, $\tau_{\mathrm{QGP}}$, 
by the following definition: At $\tau=\tau_{\mathrm{QGP}}$, the QGP fraction
 $f_{\mathrm{QGP}}$  in Eq.~(\ref{eq:frac_qgp}) at $x=y=z=0$ reaches 
  to $f_{0}$, which takes a value between 0 and 1.
 For $f_{0}=0$, the effective lifetime is defined as the time
 when QGP disappears completely, while $f_{0}=1$ corresponds to  
  the time when hadronic phase starts to appear.
 The effective lifetime of QGP is listed in Table \ref{RELAX_LIFE}(b)
  for two different impact parameters and for three different values
   of $f_{0}$. 
 
 From the comparison of $\tau_{\mathrm{Q}}$ and $\tau_{\mathrm{QGP}}$ in
 Table \ref{RELAX_LIFE}, one finds that
 the initial momentum distributions of charm and bottom quarks
 will be  changed by QGP only slightly for the weak drag force ($\gamma = 0.30$).
 On the other hand, for the strong drag force ($\gamma = 3.0$),
 both charm and bottom quarks are affected by QGP and 
 their momentum distributions would be modified substantially.  

\begin{table}
\centering
(a) \ \ \ \
\begin{tabular}{c|ccc} 
\hline\hline
\ $\gamma$ \ & \ 0.30 \ & \ 1.0 \ & \ 3.0 \ \\ 
\hline 
\ $\tau_{\rm c}$\ [fm] \ & \ 22 \ & \ 6.7 \ & \ 2.2 \ \\ 
\ $\tau_{\rm b}$\ [fm] \ & \ 72 \ & \ 21 \ & \ 7.2 \ \\ 
\hline
\end{tabular}
\ \ \ \ \ \\
\ \ \ \ \ \\
\ \ \ \ \ \\
(b) \ \ \ \ 
\begin{tabular}{c|ccc} 
\hline\hline
$f_{0}$ & 0 & 0.5 & 1 \\ 
\hline 
$\tau_{\mathrm{QGP}}^{b=3.1 {\rm fm}}$ \ [fm] & \ 9.8 \ & \ 5.9 \ & \ 4.5 \    \\ 
$\tau_{\mathrm{QGP}}^{b=5.5 {\rm fm}}$ \ [fm] & \ 8.7 \ & \ 5.2 \ & \ 4.0 \  \\ 
\hline 
\end{tabular}
\caption[RELAX_LIFE]
{\footnotesize
(a) Relaxation times of charm and bottom quarks for $\gamma=$ 0.3, 1.0, and 3.0 at 
 $T=$ 210 MeV.
(b) Lifetimes of QGP for different centralities and 
freezeout conditions.
In (a) $M_{\rm c}$ and $M_{\rm b}$ are chosen to be 1.5 GeV and 4.8 GeV, respectively.
In (b) we adopt two characteristic  impact parameters $b=$ 3.1 and 5.5 fm.
}
\label{RELAX_LIFE}
\end{table}%

%%%%%%%%%%%%%%%%%%%%%%%%%%%%%%%%%%%%%%%%%%%%%%%%%%%%%%%%%%%%%%%%%%%%%%%
\subsubsection{Initial distribution of heavy quarks}
\label{sec3b1}
%%%%%%%%%%%%%%%%%%%%%%%%%%%%%%%%%%%%%%%%%%%%%%%%%%%%%%%%%%%%%%%%%%%%%%%
 On the initial hypersurface $\tau_{0}=0.6$ fm,
initial transverse positions of heavy quarks
are distributed according to the overlap function of two 
nuclei $\rm A$ and $\rm B$ in the transverse plane $T_{\rm{AB}}(x,y)$:
\begin{eqnarray}
T_{\rm{AB}}(x,y) & = & T_{\rm A}\left(x+\frac{b}{2},y\right)
T_{\rm B}\left(x-\frac{b}{2},y\right),  \nonumber \\
T_{\rm{A(B)}}(x,y) & = & \int dz \rho_{\rm{A(B)}}(x,y,z), 
\end{eqnarray}
where $\rho_{\rm A(B)}$ is the Woods-Saxon parametrization of
nuclear density.
 The heavy quarks are assumed to stream 
freely in the longitudinal direction for $0 < \tau <\tau_0$
 and acquire the momentum rapidity $y_p=\eta_{s}$.
Thus the initial heavy quark distribution in the phase space reads
\begin{eqnarray}
\frac{dN}{d^{3}pd^{2}x_{\perp }\tau_{0}d\eta_{s}} 
&=& \frac{d\sigma^{\mathrm{HQ}}_{\rm{pp}}}{d^{3}p}T_{\rm{AB}}(x,y)\frac{
\delta(\eta_{s}-y_{p})}{\tau_{0}}  . \label{eq:HQ_ini}
\end{eqnarray}
The initial momentum spectrum
of heavy quarks $d\sigma^{\mathrm{HQ}}_{\rm{pp}}/d^{3}p$ in $p+p$ collisions
is calculated by perturbative QCD to leading order (LO)
utilizing the event generator PYTHIA 6.4 \cite{Sjostrand}.

\begin{figure}
\centering
\includegraphics[height=7cm,angle=270,origin,clip]{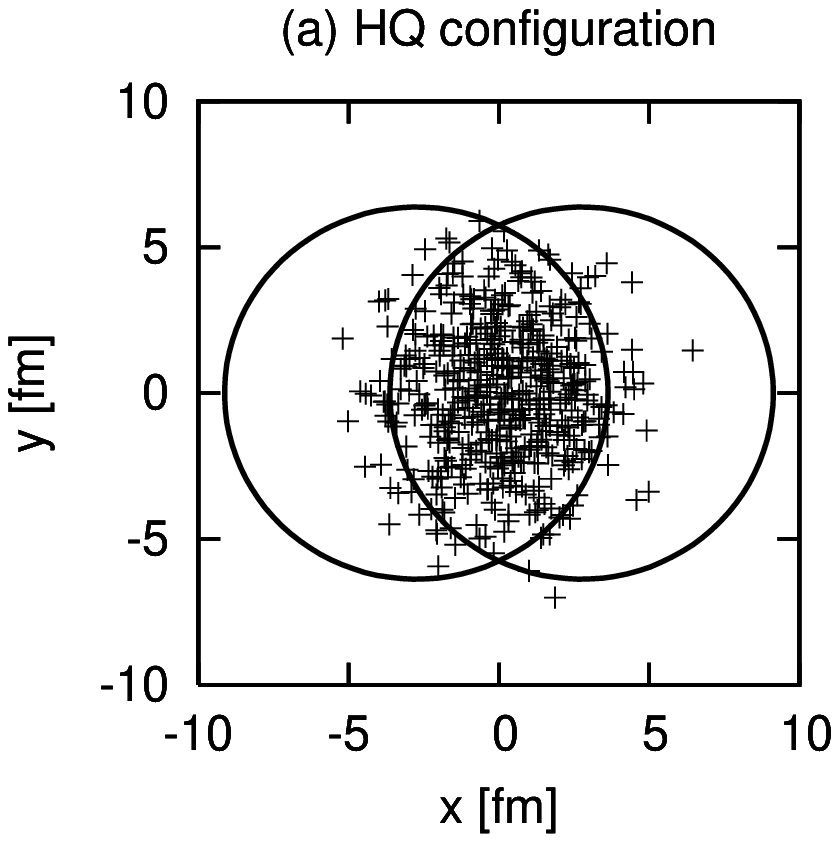}
\includegraphics[width=7cm,clip]{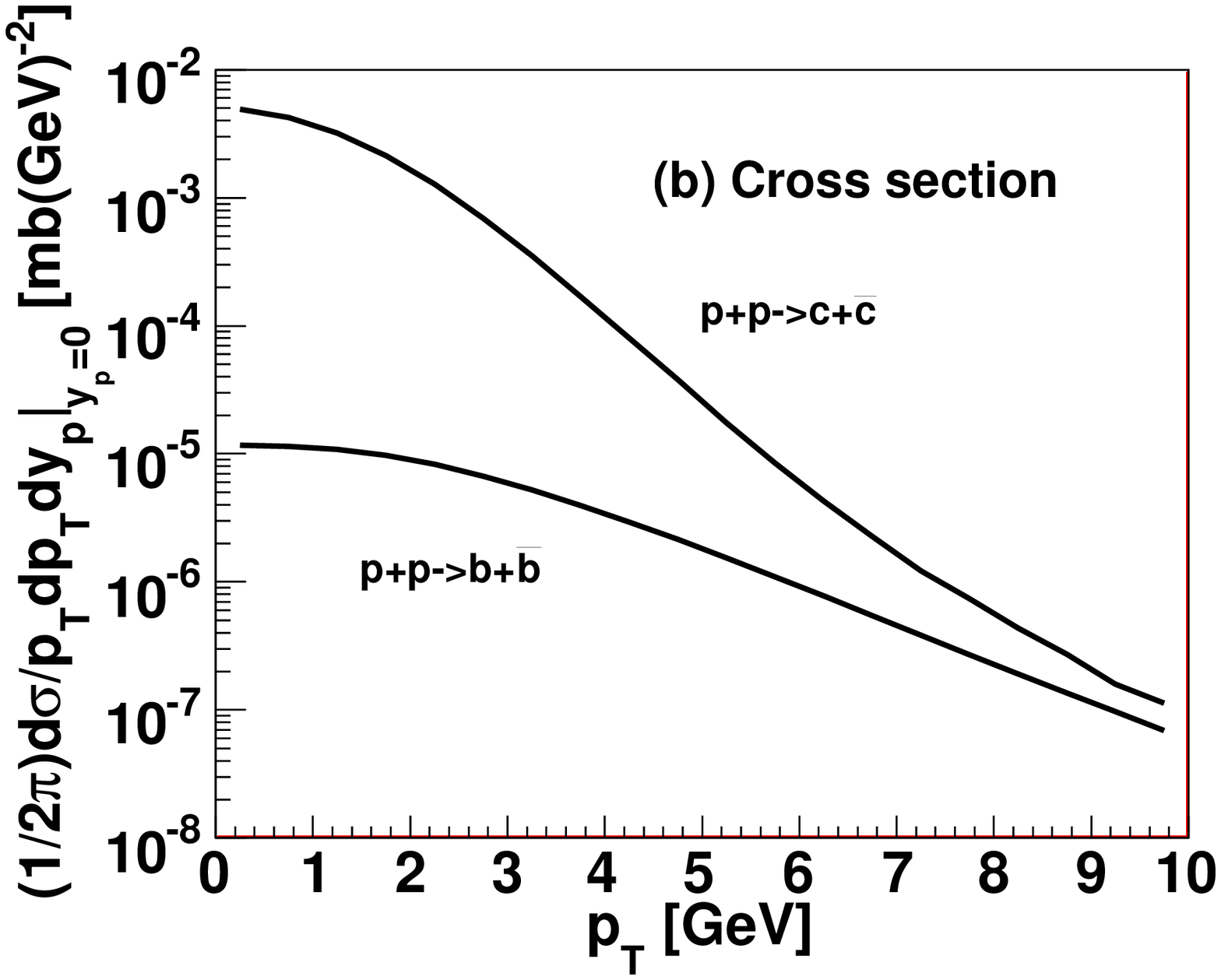}
\caption[INITIAL]
{\footnotesize 
(a) A sample of 500 heavy quarks at the initial in the transverse plane,
for the Au + Au collision with impact parameter 5.5 fm. 
(b) Invariant cross sections of charm and bottom production in $p+p$ collisions in mid-rapidity ($|y_{p}|\leq 1.0$), which is proportional to the initial momentum distribution.
}
\label{INITIAL}
\end{figure}

In Fig.~\ref{INITIAL} the initial distribution of heavy quarks
in the transverse plane and that in the momentum space at mid-rapidity
 ($|y_{p}|\leq 1$) are shown. 
The momentum distribution is normalized to the invariant cross section in $p+p$ collisions.
Note that the initial momentum distribution of the charm quark has a steeper
 slope at high $p_{\rm T}$ than that of the bottom quark.
Nuclear effects such as shadowing and Cronin effect
are not considered for simplicity.
%\begin{widetext}
%\end{widetext}

\begin{figure}
\centering
\includegraphics[width=7cm,clip]{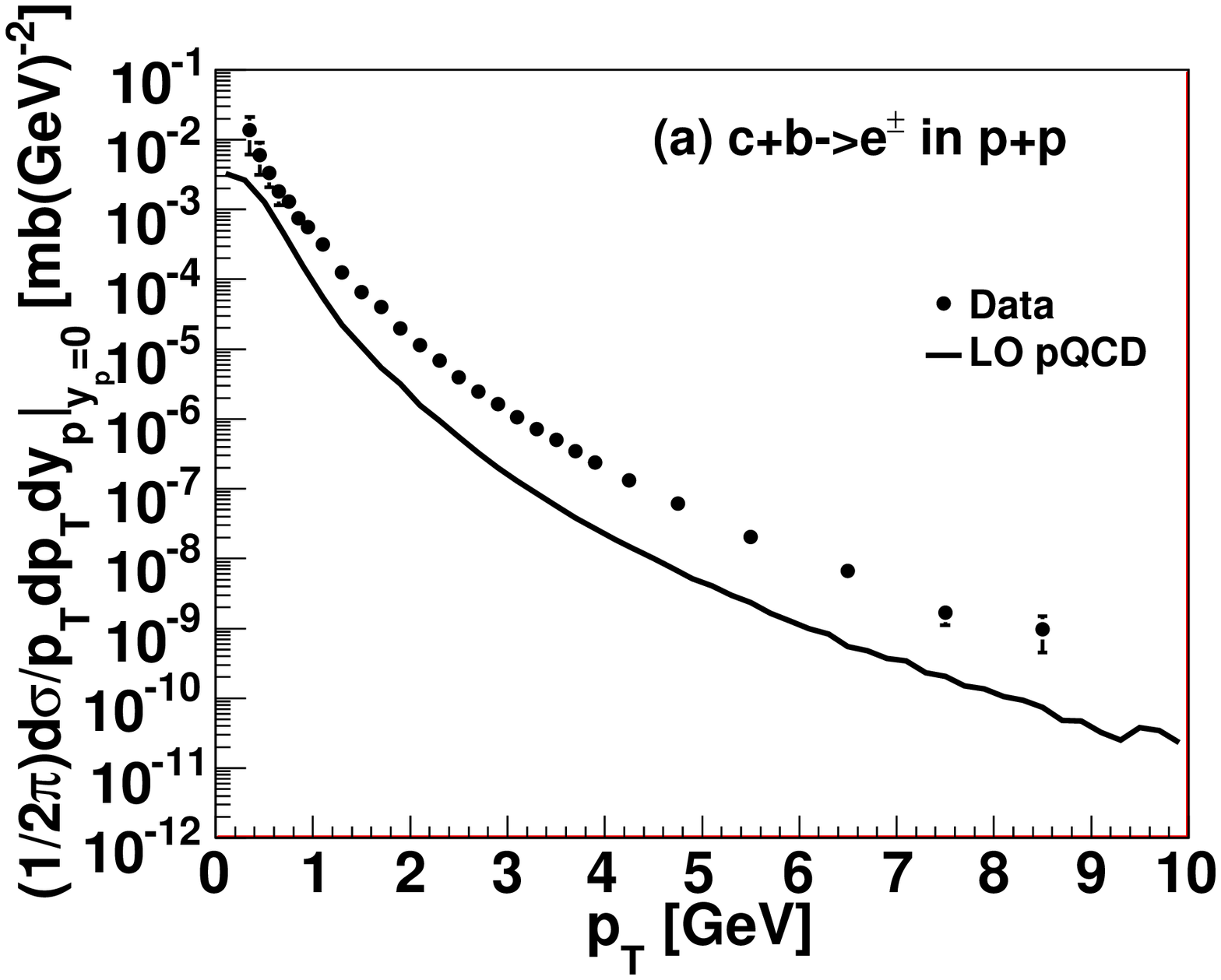}
\includegraphics[height=7cm,angle=270,origin,clip]{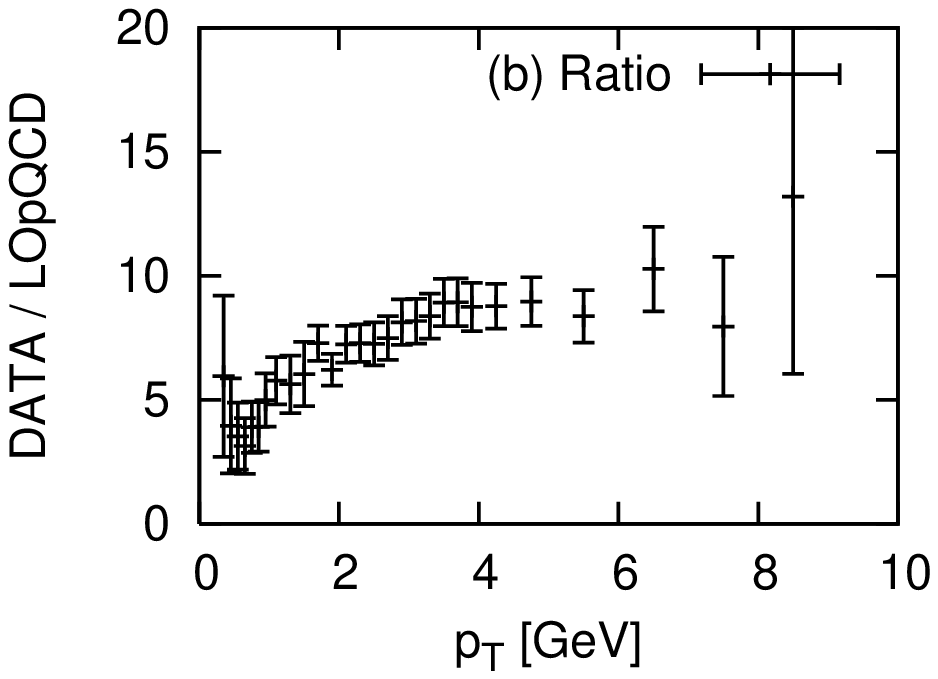}
\caption[CROSS]
{\footnotesize 
(a) Experimental cross section for electron production in $p+p$ collision
  at mid-rapidity \cite{Adare:pp} and the leading order pQCD result by PYTHIA. 
(b)  The ratio of the experimental data and the LO result.
 Theoretical calculations are performed
  at $|y_{p}|\leq 0.35$  and then properly normalized to obtain the cross section.
}
\label{CROSS}
\end{figure}

In  Fig.~\ref{CROSS}(a), we show the differential cross section of electrons
from heavy quarks in $p+p$ collisions obtained by PYTHIA (the LO perturbative
 QCD).
  Theoretical cross section underestimates
   the experimental value by a factor $5$-$10$
  as shown in  Fig.~\ref{CROSS}(b).
 The discrepancy is known to become smaller by taking into account higher orders
 beyond LO \cite{Adare:pp,Cacciari}.  Since we are  mainly interested in the 
 the ``shape" of the $p_{\rm T}$ distribution above 3 GeV 
in this paper, we adopt the LO result for simplicity in spite of  the problem 
 of absolute magnitude in the LO calculation.

%%%%%%%%%%%%%%%%%%%%%%%%%%%%%%%%%%%%%%%%%%%%%%%%%%%%%%%%%%%%%%%%%%%%%%%
\subsubsection{Simulation of the Brownian motion}
\label{sec3b2}
%%%%%%%%%%%%%%%%%%%%%%%%%%%%%%%%%%%%%%%%%%%%%%%%%%%%%%%%%%%%%%%%%%%%%%%

The Langevin equation of a heavy quark 
is defined in the local rest frame of QGP.
 The information of local flow velocity and local 
temperature at the position of the heavy quark is supplied from
 the relativistic hydrodynamics.
The algorithm of such Langevin simulation 
is summarized as follows. \\

\noindent
(i) 
Start from a sample of heavy quark at a position and a momentum
according to the initial phase space distribution 
given in Eq.~(\ref{eq:HQ_ini}).\\

\noindent
(ii) 
Given the phase space location $(p^{\mu},x^{\mu})$ in the laboratory frame,
obtain the information of the local flow velocity $u^{\mu}(x)$
and local temperature $T(x)$ 
 from the output of the hydrodynamic simulation.  \\

\noindent
(iii-a) Coordinate step:
 Make one discrete step for the heavy quark
  in the configuration space according to Eq.~(\ref{eq:dx})
by using discrete proper-time step $\Delta s= (M/E) \Delta t$:
\begin{eqnarray}
\Delta x^{\nu}(s)=\frac{p^{\nu}}{M}\Delta s .
\end{eqnarray}

\noindent
(iii-b) Momentum step: Move to the rest frame of the fluid element 
by the Lorentz transformation, $p\rightarrow k$.
 Make one discrete step for the heavy quark
  in momentum according to Eq.~(\ref{eq:dp})
 using $\Delta s$:
\begin{eqnarray}
\Delta\vec k(s) &=& -\gamma \frac{T^{2}E(k)}{M^{2}}\vec k\Delta s + \vec \xi (s), \\
\langle \xi_{i}(s) \xi_{j}(s') \rangle &=& \delta_{ij} \delta_{ss'} 
\frac{2\gamma T^{3}}{M^{2}}E(E+T)\Delta s .
\end{eqnarray}
Then, move back to the laboratory frame by inverse Lorentz transformation
($k+\Delta k\rightarrow p'$).\\

\noindent
(iv)
Repeat the steps (ii) and (iii) until the volume fraction of QGP in the mixed phase ($f_{\rm QGP}$) reaches $f_{0}$.\\

Several comments are in order here about this procedure.
\begin{itemize}
  \item
   We use the proper-time step $\Delta s$ 
 instead of the ordinary time step $\Delta t$ in our simulation,
  simply because the former
 is  a Lorentz scalar and thus easy to handle in going back and forth between
  the laboratory frame and the fluid rest frame.
We choose $\Delta s=0.01$ fm in our simulation,
 which is much shorter than the relaxation time
of the drag force parameter adopted in this paper.
 \item
 Owing to the It\^{o} discretization scheme, the momentum step 
in (iii-b) can be performed only by using the information 
of flow and temperature at the current position of 
the heavy quark in the phase space.
\item
It is not clear whether we should stop the heavy quark
 diffusion at the point when the mixed phase starts to appear
  or at the point when the mixed phase disappears. 
 We consider this uncertainty as a 
 systematic error and consider the three cases as 
  shown in Table \ref{RELAX_LIFE}(b), namely
 $f_{0}=0, 0.5$, and 1.
\end{itemize}

%%%%%%%%%%%%%%%%%%%%%%%%%%%%%%%%%%%%%%%%%%%%%%%%%%%%%%%%%%%%%%%%%%%%%%%
\subsubsection{Freezeout and decay}
\label{sec3b3}
%%%%%%%%%%%%%%%%%%%%%%%%%%%%%%%%%%%%%%%%%%%%%%%%%%%%%%%%%%%%%%%%%%%%%%%
Once the local temperature around the charm(bottom) quark becomes lower than 
$T_{\rm c}$,
it hadronizes into $D$ ($B$) mesons.
 Since we need to calculate single electron spectra
 from the heavy quarks, we focus
on the following semileptonic decays: 
$D\rightarrow e$ for $D$ decay, 
$B\rightarrow e$ for primary $B$ decay,
and
$B\rightarrow D\rightarrow e$ for secondary $B$ decay. 
The hadronization of heavy quarks and the decay of heavy mesons
 are calculated by using PYTHIA 6.4 \cite{Sjostrand}.  
 Since we employ independent fragmentation given by PYTHIA,
 the effect of quark recombination
 to form $D$ or $B$ mesons is not taken into account.
 Such simplification would be more reasonable 
for heavy quarks with higher transverse momentum.
Therefore, it is the high $p_{\rm T}$ region (e.g. above 3 GeV) 
that is  suitable to compare our results
with the experimental data.

%%%%%%%%%%%%%%%%%%%%%%%%%%%%%%%%%%%%%%%%%%%%%%%%%%%%%%%%%%%%%%%%%%%%%%%
\subsection{Observables}
\label{sec3c}
%%%%%%%%%%%%%%%%%%%%%%%%%%%%%%%%%%%%%%%%%%%%%%%%%%%%%%%%%%%%%%%%%%%%%%%
 
 Medium modification factor $R_{\rm{AA}}$ for single electrons
 is defined by
\begin{eqnarray}
R_{\rm{AA}}(p_{\rm T})&=&\frac{1}{N_{\rm{coll}}}\frac{dN_{\rm{A+A}}/dp_{\rm T}}{dN_{p+p}/dp_{\rm T}},  
\label{eq:RAA}
\end{eqnarray}
where $N_{\rm{coll}}$ is the number of binary collisions calculated from 
the Glauber model.
Since the initial heavy quark distribution is assumed to be without nuclear effects and to scale as
 $N_{\rm{coll}}$ in our calculation, the deviation of $R_{\rm{AA}}$
from unity is solely attributed to the heavy quark diffusion in the hot medium.
The elliptic flow for single electrons is defined by
\begin{eqnarray}
v_{2}(p_{\rm T})&=&\frac{\int d\phi \frac{d^{2}N_{\rm{A+A}}}{dp_{\rm T}d\phi}\cos2\phi}
{\int d\phi \frac{d^{2}N_{\rm{A+A}}}{dp_{\rm T}d\phi}}
\ = \ \langle \cos2\phi \rangle. 
\end{eqnarray}
This quantity indicates 
 how much momentum anisotropy around the collision axis is given to the heavy quarks 
from the background medium.

%%%%%%%%%%%%%%%%%%%%%%%%%%%%%%%%%%%%%%%%%%%%%%%%%%%%%%%%%%%%%%%%%%%%%%%
\section{Numerical Results}
\label{sec4}
%%%%%%%%%%%%%%%%%%%%%%%%%%%%%%%%%%%%%%%%%%%%%%%%%%%%%%%%%%%%%%%%%%%%%%%
 
Before showing the numerical results in detail, let us first summarize the basic 
parameters of our simulation.
(1)
The dimensionless drag coefficient $\gamma$ is a 
 parameter to control the diffusion of heavy quarks in QGP.  
 We take three characteristic values, $\gamma$ = 0.3, 1.0, and 3.0 corresponding
  to weak, intermediate, and strong coupling, respectively.
(2)
The impact parameter $b$ controls the volume and the lifetime of QGP. Thus it
 affects indirectly the heavy quark spectra at their freezeout and 
 the single electron spectra.
In all of the figures below except for Fig.~\ref{SPECTRUM_EP_RAA_EX},
 $b$ is taken to be 5.5 fm (10-20\% centrality).
(3)
The criterion of stopping the heavy quark diffusion in the mixed phase
 is given by $f_{0}$ which takes a value between 0 and 1.
 Its dependence on the final results 
 is considered to be a  systematic error of our calculation.
In all of the figures except
 for Fig.~\ref{SPECTRUM_EP_RAA_EX}, we show the results at the 
  central value, $f_{0}=0.5$.

%%%%%%%%%%%%%%%%%%%%%%%%%%%%%%%%%%%%%%%%%%%%%%%%%%%%%%%%%%%%%%%%%%%%%%%
\subsection{Heavy quark spectra}
\label{sec4a}
%%%%%%%%%%%%%%%%%%%%%%%%%%%%%%%%%%%%%%%%%%%%%%%%%%%%%%%%%%%%%%%%%%%%%%%

%%%%%%%%%%%%%%%%%%%%%%%%%%%%%%%%%%%%%%%%%%%%%%%%%%%%%%%%%%%%%%%%%%%%%%%
\subsubsection{Profile of heavy quark diffusion}
\label{sec4a1}
%%%%%%%%%%%%%%%%%%%%%%%%%%%%%%%%%%%%%%%%%%%%%%%%%%%%%%%%%%%%%%%%%%%%%%%

To estimate how long a heavy quark stays in the QGP region in terms of local
 fluid proper
 time, we define the ``stay time" as 
\begin{eqnarray}
\label{ave_time}
t_{\rm S}&\equiv&\sum_{\rm{steps}} \Delta t|_{\mathrm{FRF}}=\sum_{\rm{steps}} \Delta s (E/M)|_{\mathrm{FRF}}\nonumber \\
&=&\sum_{\rm{steps}} \Delta s (p\cdot u/M)|_{\mathrm{LF}},
\end{eqnarray}
where FRF and LF imply the fluid rest frame and laboratory frame, respectively.
By averaging over the heavy quarks starting
initially with $p_{\rm T}^{\rm in}$ and ending in mid-rapidity ($|y_{p}|\leq$ 1.0) at 
their freezeout,
 we obtain the average stay time $\langle t_{\rm S}\rangle$.  

\begin{figure}
\centering
\includegraphics[width=7cm,clip]{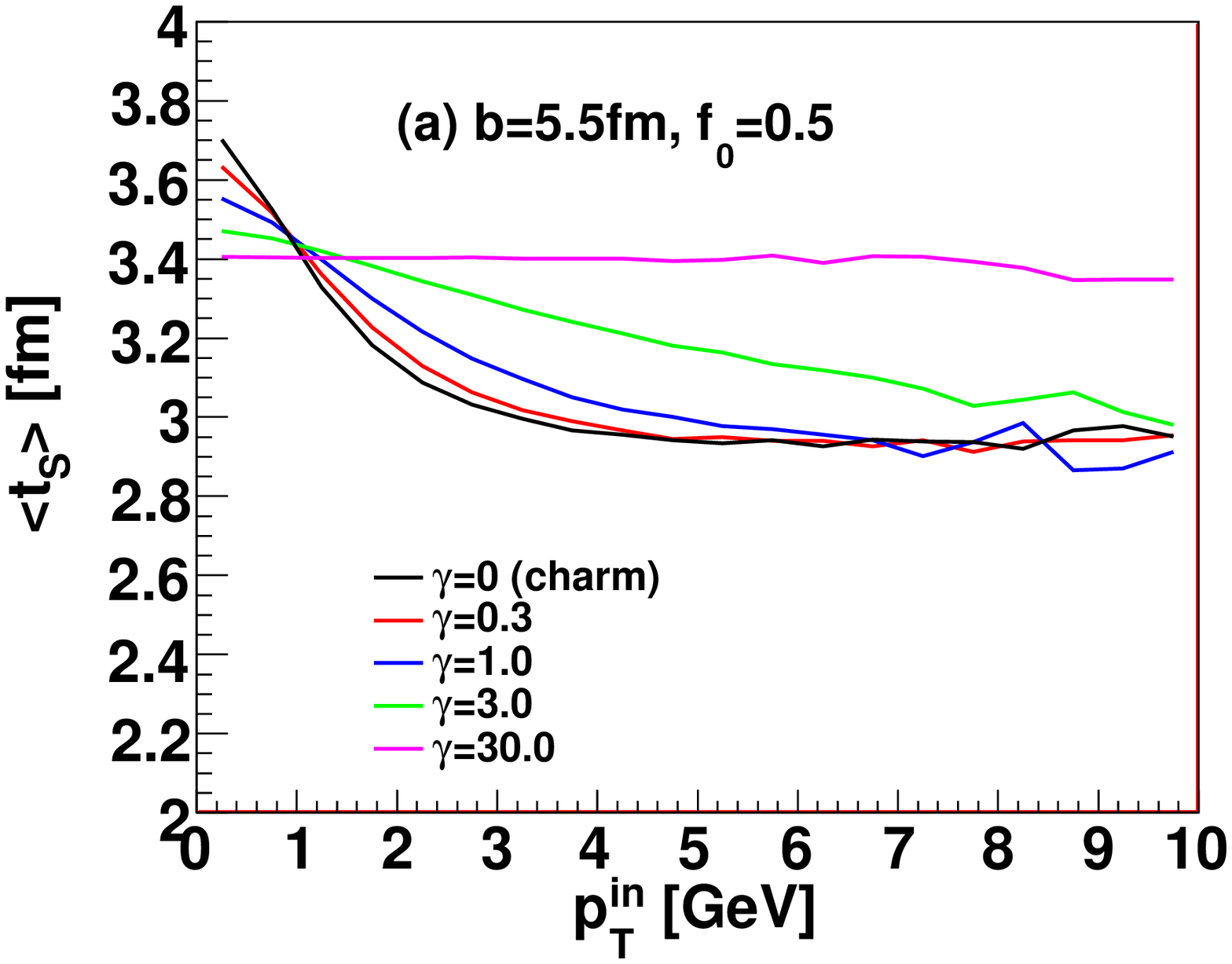}
\includegraphics[width=7cm,clip]{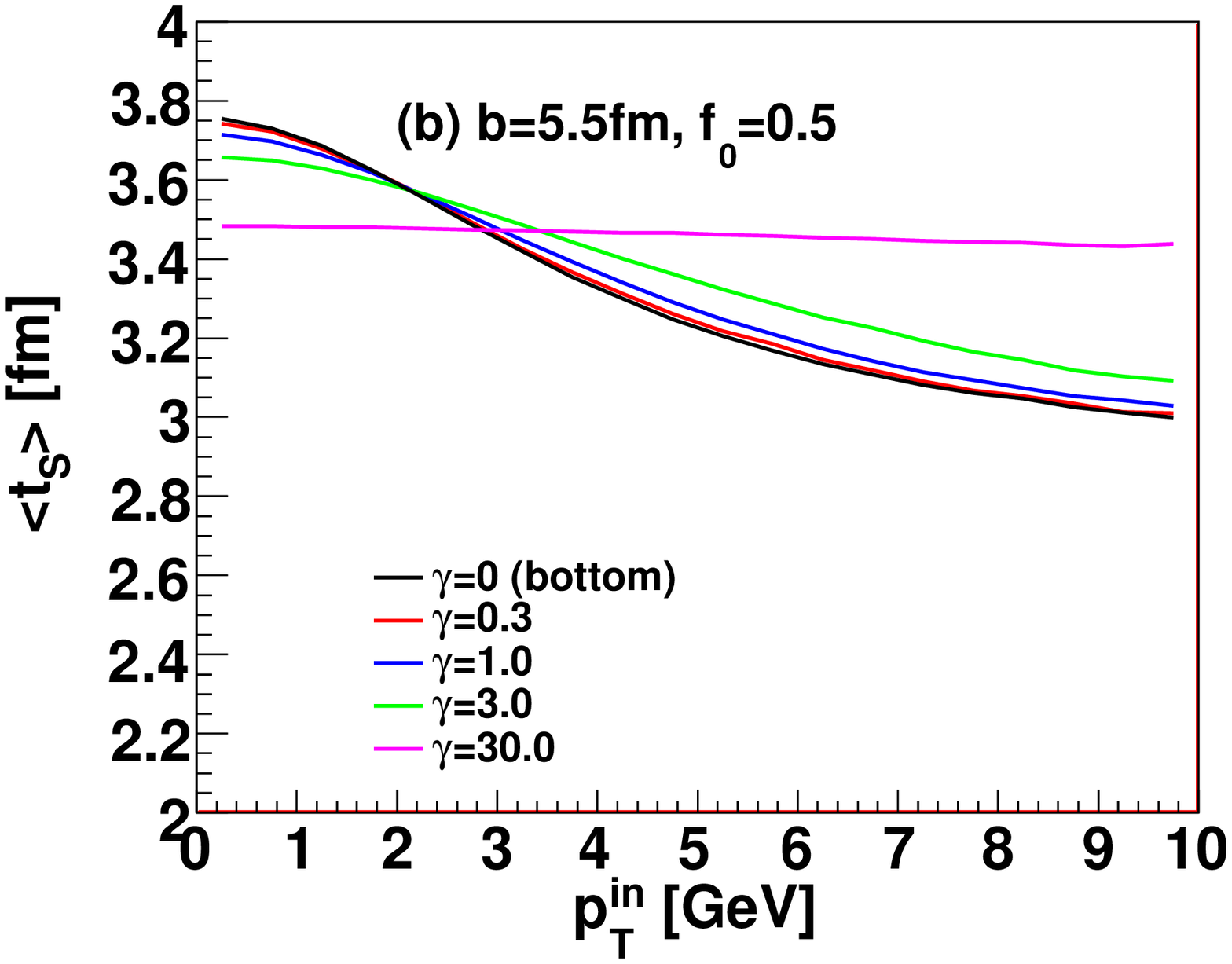}
\caption[PROFILE_TIME]
{\footnotesize (Color online)
 The averaged
 stay time $\langle t_{\rm S}\rangle$ of (a) charm quarks and (b) bottom quarks with the drag coefficient $\gamma=$ 0, 0.3, 1.0, 3.0, and 30.0 at mid-rapidity ($|y_{p}|\leq1.0$).
The impact parameter is chosen to be 5.5 fm in Au+Au collisions. 
For freezeout condition, $f_{0}=$ 0.5 is adopted.
}
\label{PROFILE_TIME}
\end{figure}

Shown in Fig.~\ref{PROFILE_TIME}
 is the averaged stay time of heavy quarks as a function of their
  initial transverse momentum.  The diffusion coefficient  is taken to be
 $\gamma$ = 0, 0.3, 1.0, 3.0, and 30.0.
 Here $\gamma$ = 0 corresponds to the free streaming. On the other hand,
 $\gamma=30.0$ corresponds to the extremely strong coupling
  where  the relaxation times at typical temperature 210 MeV are 0.22 fm for charm and 0.72 fm for bottom: The initial information on $p_{\rm T}$ 
  is completely lost after a few fm of diffusion in this case.

  The figure shows that, for heavy quarks with large initial velocity
 compared to the fluid velocity
  ($p^{c,\rm in}_{\rm T}>$ 1 GeV, $p^{b,\rm in}_{\rm T}>$ 3 GeV),
  the stay time becomes shorter for higher $p_{\rm T}$ 
  because they  get out of the medium in shorter times.
  Also, as the drag force becomes stronger, the stay time becomes longer as expected.
  As for the heavy quarks with small initial velocity
   ($p^{c,\rm in}_{\rm T}<$ 1 GeV, $p^{b,\rm in}_{\rm T}<$ 3 GeV),
  the stronger the drag force, the shorter the stay time, since
the drag force from the background fluid accelerates them more strongly.
%The crossing momentum may be determined mainly by the flow profile of 
%the background fluid.

\begin{figure}
\centering
\includegraphics[width=7cm,clip]{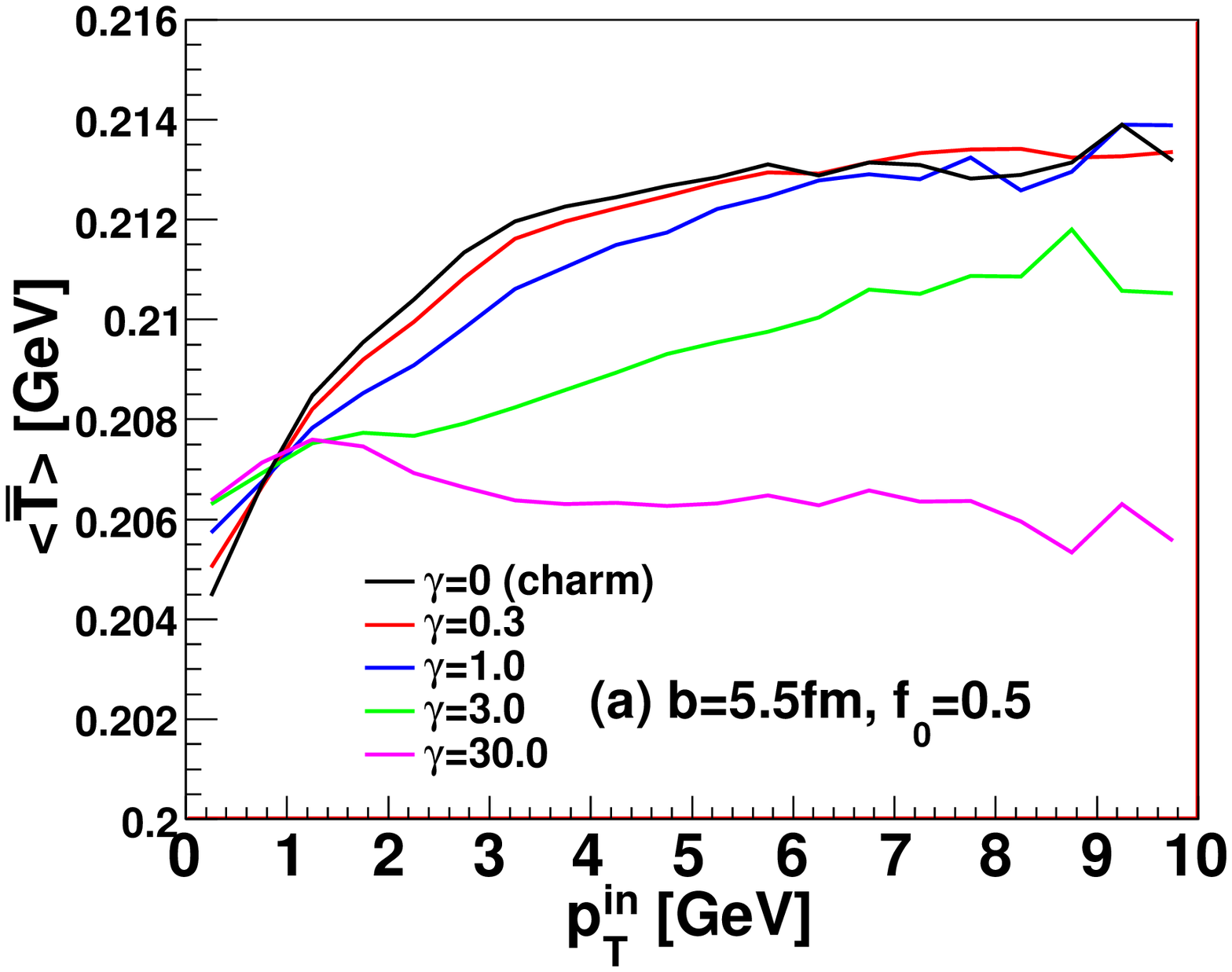}
\includegraphics[width=7cm,clip]{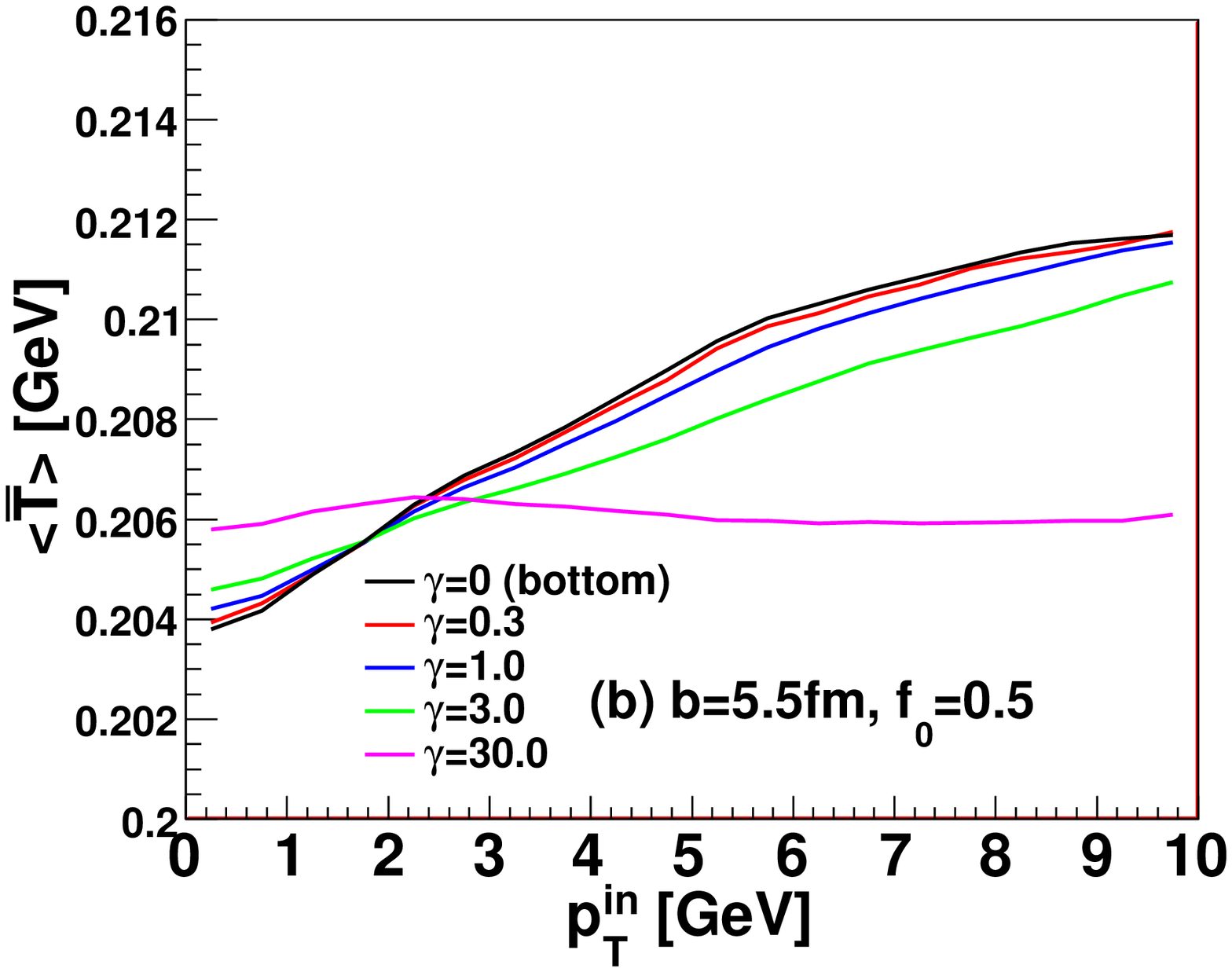}
\caption[PROFILE_TEMP]
{\footnotesize  (Color online)
The averaged temperature $\langle \bar{T} \rangle$ of (a) charm quarks and (b) bottom quarks with the 
drag coefficient $\gamma=$ 0, 0.3, 1.0, 3.0, and 30.0 in mid-rapidity ($|y_{p}|\leq1.0$).
The collision geometry and the freezeout condition are the same with those in 
Fig.~\ref{PROFILE_TIME}.
The fluctuation in high $p_{\rm T}$ in (a) is due to statistical errors of our
 simulation.
}
\label{PROFILE_TEMP}
\end{figure}

Next we define the averaged temperature
 for the heavy quarks experienced during their stay in the QGP fluid:
\begin{eqnarray}
\label{ave_temp}
\bar T\equiv(1/t_{\rm S})\sum_{\rm{steps}} T(x)\Delta t|_{\mathrm{FRF}} .
\end{eqnarray}
By averaging over the heavy quarks starting
initially with $p_{\rm T}^{\rm in}$
 and ending in mid-rapidity ($|y_{p}|\leq$ 1.0) at freezeout,
 we obtain the averaged temperature $\langle \bar T\rangle$
  shown in Fig.~\ref{PROFILE_TEMP}.

   The figure shows that, for heavy quarks with large initial velocity
  compared to the fluid velocity ($p^{c,\rm in}_{\rm T}>$ 1 GeV, $p^{b,\rm in}_{\rm T}>$ 3 GeV),
  the averaged temperature becomes higher for higher $p_{\rm T}^{\rm in}$ 
  because they feel only the initial high temperature region before
  getting out of QGP.
  Also, as the drag force becomes stronger, the stay time becomes longer 
  and averaged temperature becomes smaller.
  As for the heavy quarks with small initial velocity
   ($p^{c,\rm in}_{\rm T}<$ 1 GeV, $p^{b,\rm in}_{\rm T}<$ 3 GeV),
  the stronger the drag force, the higher the averaged temperature, since
they are strongly accelerated and quickly pass the low temperature region.
%The crossing momentum is determined by the flow profile and the temperature profile of %the background fluid.
 It turns out that the average temperature lies between 200 MeV and 220 MeV 
  in the wide range of $p_{\rm T}^{\rm in}$ and $\gamma$; this is
   the reason why we adopted the typical temperature 210 MeV in Sec.~\ref{sec3}.

\begin{figure}
\centering
\includegraphics[width=7cm,clip]{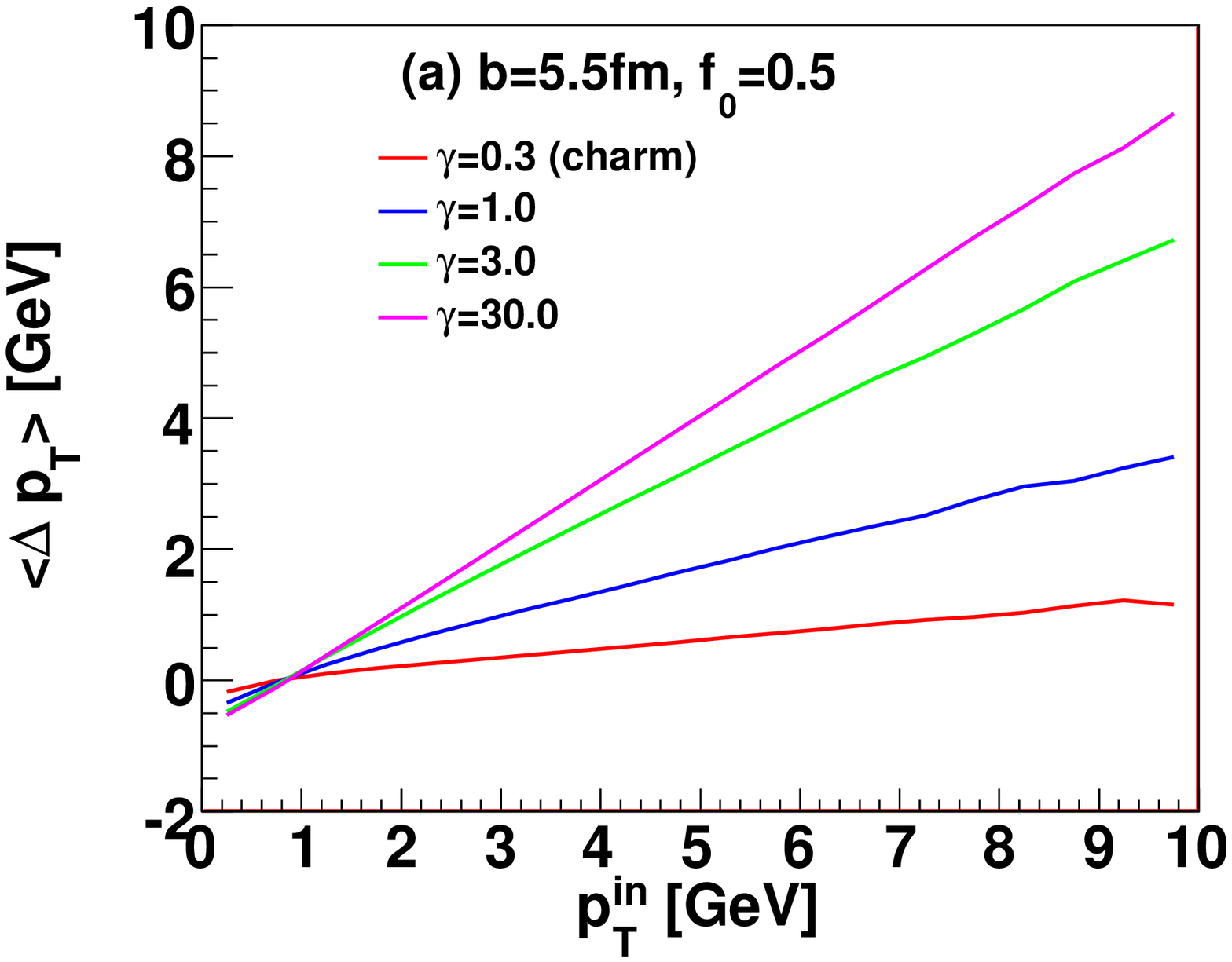}
\includegraphics[width=7cm,clip]{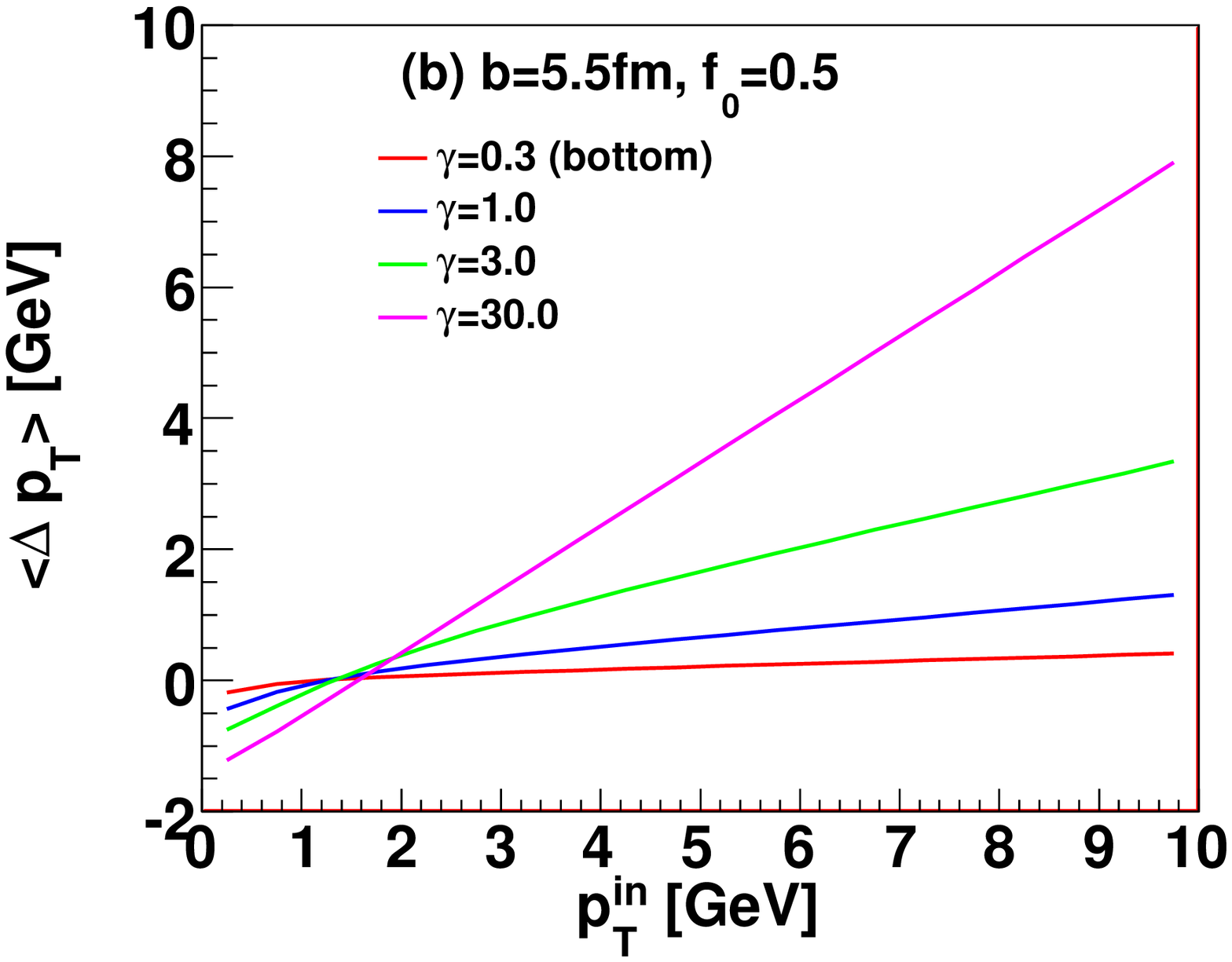}
\caption[PROFILE_PTLOSS]
{\footnotesize  (Color online)
The averaged momentum loss 
$\langle \Delta p_{\rm T}\rangle$ of (a) charm quarks and (b) bottom quarks with drag coefficients $\gamma=$ 0.3, 1.0, 3.0, and 30.0 in mid-rapidity ($|y_{p}|\leq 1.0$).
The collision geometry and the freezeout condition are the same with those in Fig.~\ref{PROFILE_TIME}.
}
\label{PROFILE_PTLOSS}
\end{figure}

Finally, let us define the 
the transverse momentum loss (momentum loss for short):
\begin{eqnarray}
\label{ave_ptloss}
\Delta p_{\rm T}=p_{\rm T}^{\mathrm{in}}-p_{\rm T}^{\mathrm{out}},
\end{eqnarray}
where $p_{\rm T}^{\rm out}$ is the transverse momentum at the time of
 the freezeout of the heavy quark.
By averaging over the heavy quarks starting
initially with $p_{\rm T}^{\rm in}$
 and ending in mid-rapidity ($|y_{p}|\leq$ 1.0) at freezeout,
 we obtain the averaged momentum loss $\langle \Delta p_{\rm T}\rangle$
  as shown in Fig.~\ref{PROFILE_PTLOSS}.

 For heavy quarks with larger initial $p_{\rm T}^{\rm in}$,
  the momentum loss per unit time (dynamical effect)
 is larger as seen in Eqs.~(\ref{eq:dp}) and (\ref{eq:G-param}) while
  the average stay time (kinematical effect) is shorter. Therefore, there are two 
  competing effects  in the net momentum loss:
 In Fig.~\ref{PROFILE_PTLOSS}, we find that larger initial momentum 
 leads to larger  momentum loss, so that the 
  dynamical effect wins over the kinematical effect. 
 As for the dependence on drag coefficient, both dynamical and kinematical
  effects act additively for heavy quarks with large initial momentum ($p^{c,\rm in}_{\rm T}>$ 1 GeV, $p^{b,\rm in}_{\rm T}>$ 1.5 GeV))
 and the momentum loss is enhanced by increasing $\gamma$.
For the heavy quarks with small initial velocity ($p^{c,\rm in}_{\rm T}<$ 1 GeV, $p^{b,\rm in}_{\rm T}<$ 1.5 GeV),
these two effects compete but we find in Fig.~\ref{PROFILE_PTLOSS} that the dynamical effect seems to win, namely that the stronger the drag force, 
the larger the momentum gain by the acceleration from the fluid. 
 Note here that, for the extreme case $\gamma=$ 30.0,
  we have almost a linear increase of
   $\langle\Delta p_{\rm T}\rangle=p_{\rm T}^{\rm in}- 
   p_{\rm T}^{\rm out}$ as a function of $p_{\rm T}^{\rm in}$.
     This is simply because the heavy quarks are 
     thermalized and $p_{\rm T}^{\rm out}$ is  
     almost independent of $p_{\rm T}^{\rm in}$. 
%The crossing momentum may be determined mainly by the temperature and the flow of the background fluid at freezeout.

%%%%%%%%%%%%%%%%%%%%%%%%%%%%%%%%%%%%%%%%%%%%%%%%%%%%%%%%%%%%%%%%%%%%%%%
\subsubsection{Nuclear modification factor $R_{\rm{AA}}^Q$}
\label{sec4a2}
%%%%%%%%%%%%%%%%%%%%%%%%%%%%%%%%%%%%%%%%%%%%%%%%%%%%%%%%%%%%%%%%%%%%%%%
\begin{figure}
\centering
\includegraphics[width=7cm,clip]{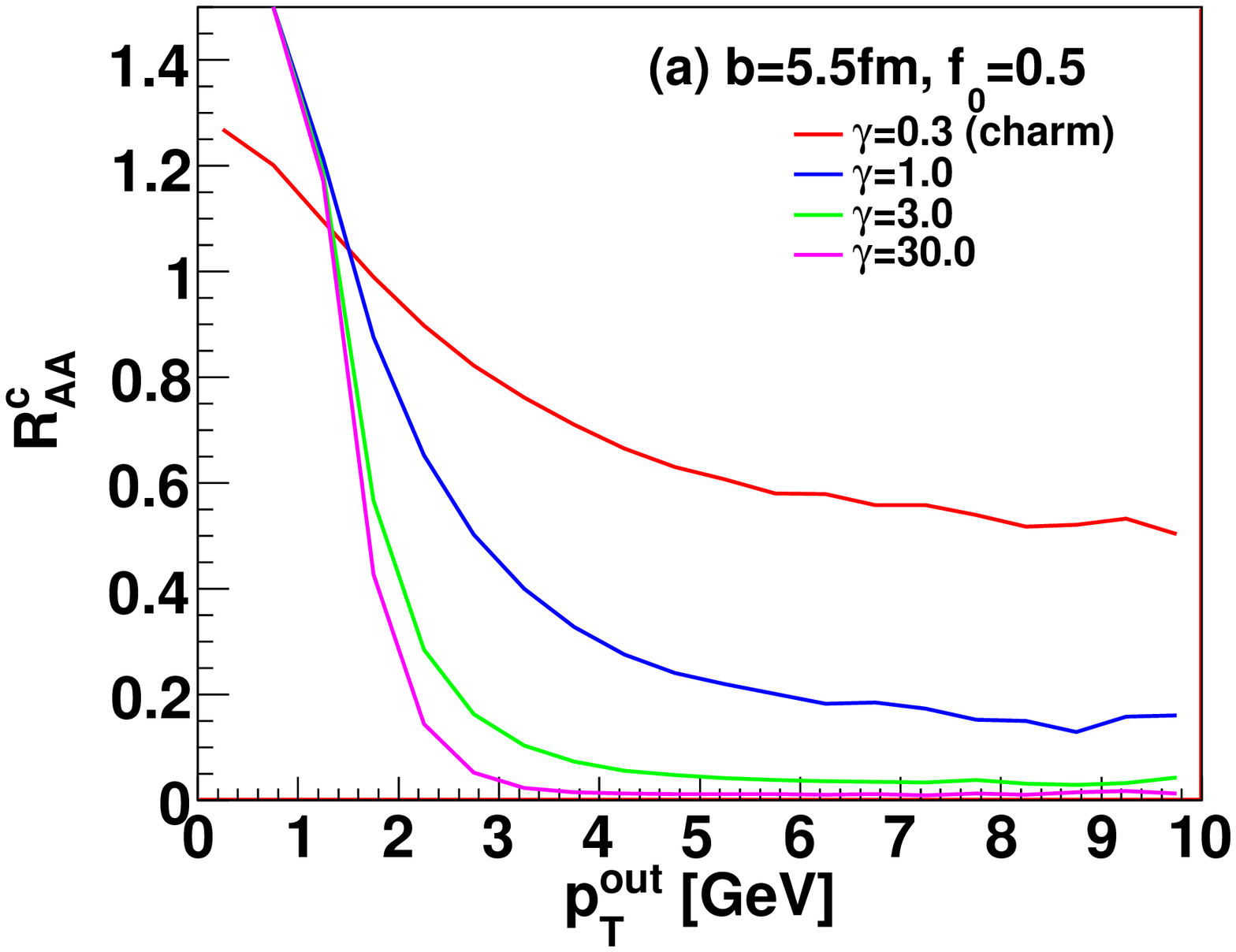}
\includegraphics[width=7cm,clip]{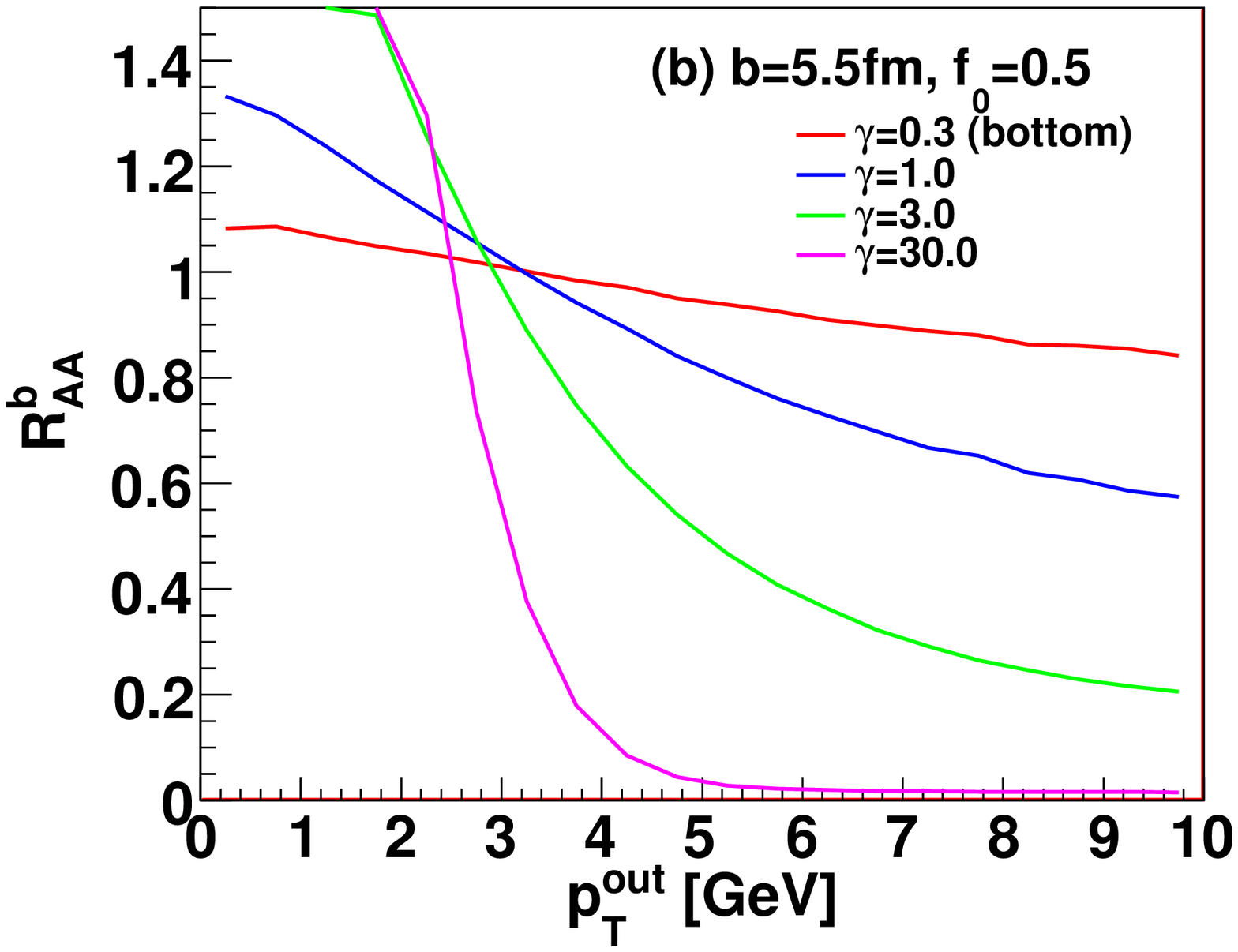}
\caption[SPECTRUM_HQ_RAA]
{\footnotesize  (Color online)
$R_{\rm{AA}}^Q$ of (a) charm quarks and (b) bottom quarks with drag coefficients $\gamma=$ 0.3, 1.0, 3.0, and 30.0 at mid-rapidity ($|y_{p}|\leq1.0$). 
For collision geometry, we choose the impact parameter 5.5 fm in Au+Au collisions. 
For freezeout condition, the $f_{0}=$ 0.5 is adopted.
}
\label{SPECTRUM_HQ_RAA}
\end{figure}

Let us define $R_{\rm{AA}}^Q$ ($Q=c, b$) for heavy quarks 
  by replacing the number of electrons 
  $N_{p+p}$ ($N_{\rm{A+A}}$) in Eq.~(\ref{eq:RAA})
    by the number of heavy quarks at the freezeout 
  $N_{p+p}^Q$ ($N_{\rm{A+A}}^Q$).  This is a theoretical 
 quantity not directly accessible in experiment, but it is
 useful to examine the behavior of heavy quarks  without 
 the kinematical complication due to their
 semileptonic decays to electrons.
  
 Shown in  Fig.~\ref{SPECTRUM_HQ_RAA} are  $R_{\rm{AA}}^Q$ for charm and bottom
   in the mid-rapidity at impact parameter 5.5 fm as a function of $p_{\rm T}^{\rm out}$.
There are two key factors which determine $R_{\rm{AA}}^Q$;  
the momentum loss of heavy quarks and the initial distribution of heavy quarks. 
Starting from the initial distribution, the high momentum
 quarks loose energy due to drag force and are shifted 
  to the low $p_{\rm T}^{\rm out}$ region.
   Therefore, $R_{\rm{AA}}^Q$ is suppressed (enhanced) 
   at high (low) $p_{\rm T}^{\rm out}$.  This tendency is prominent
   for large drag force as expected.
   Also, the suppression at high $p_{\rm T}^{\rm out}$ is larger
    for the charm if we adopt the same $\gamma$.  This is because
     the actual drag coefficient is $\gamma T^2/M_{Q}$ so that
     the quark with smaller mass is  affected more by the drag force.

%%%%%%%%%%%%%%%%%%%%%%%%%%%%%%%%%%%%%%%%%%%%%%%%%%%%%%%%%%%%%%%%%%%%%%%
\subsubsection{Elliptic flow $v_{2}^Q$}
\label{sec4a3}
%%%%%%%%%%%%%%%%%%%%%%%%%%%%%%%%%%%%%%%%%%%%%%%%%%%%%%%%%%%%%%%%%%%%%%%
%\begin{widetext}
\begin{figure}
\centering
\includegraphics[width=7cm,clip]{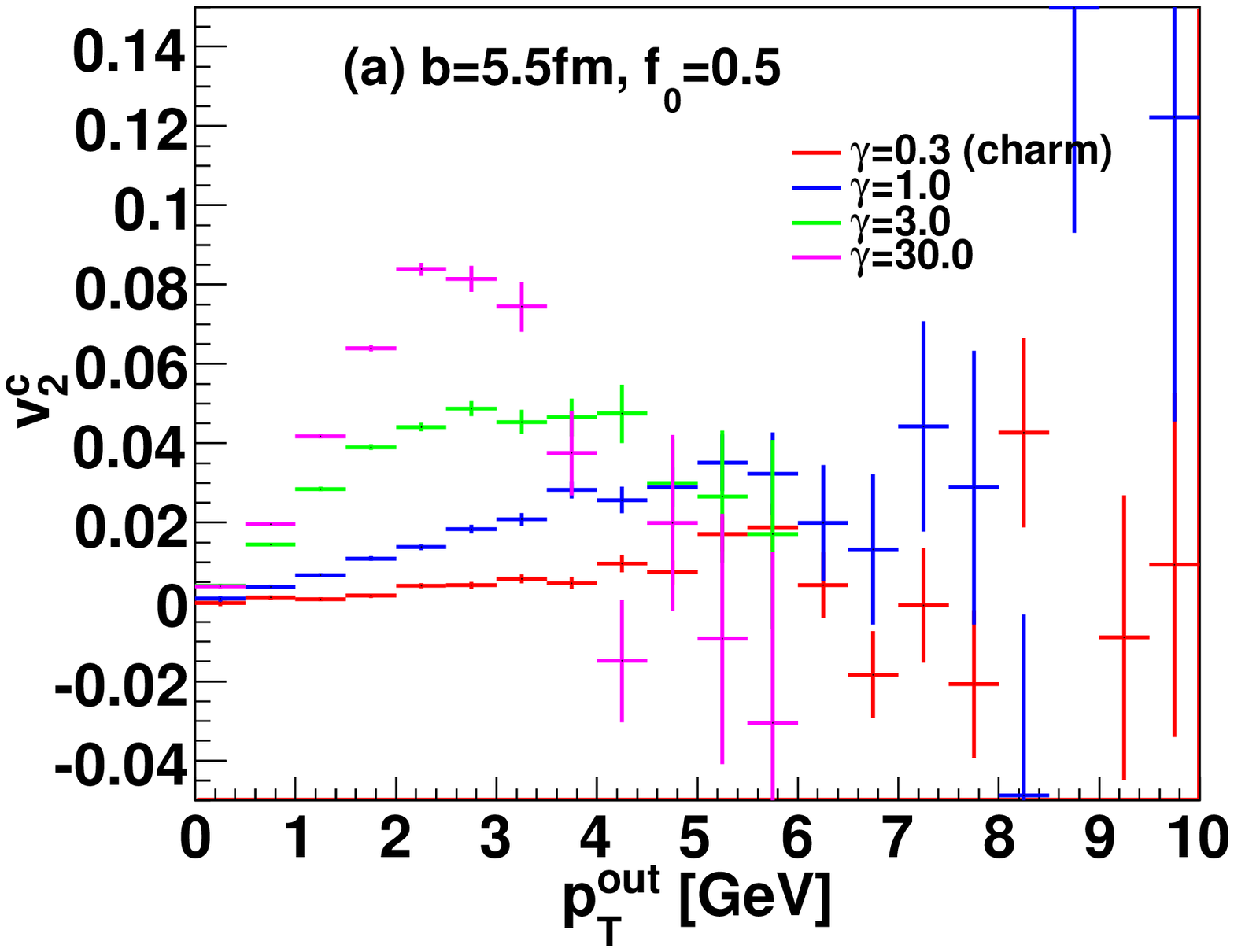}
\includegraphics[width=7cm,clip]{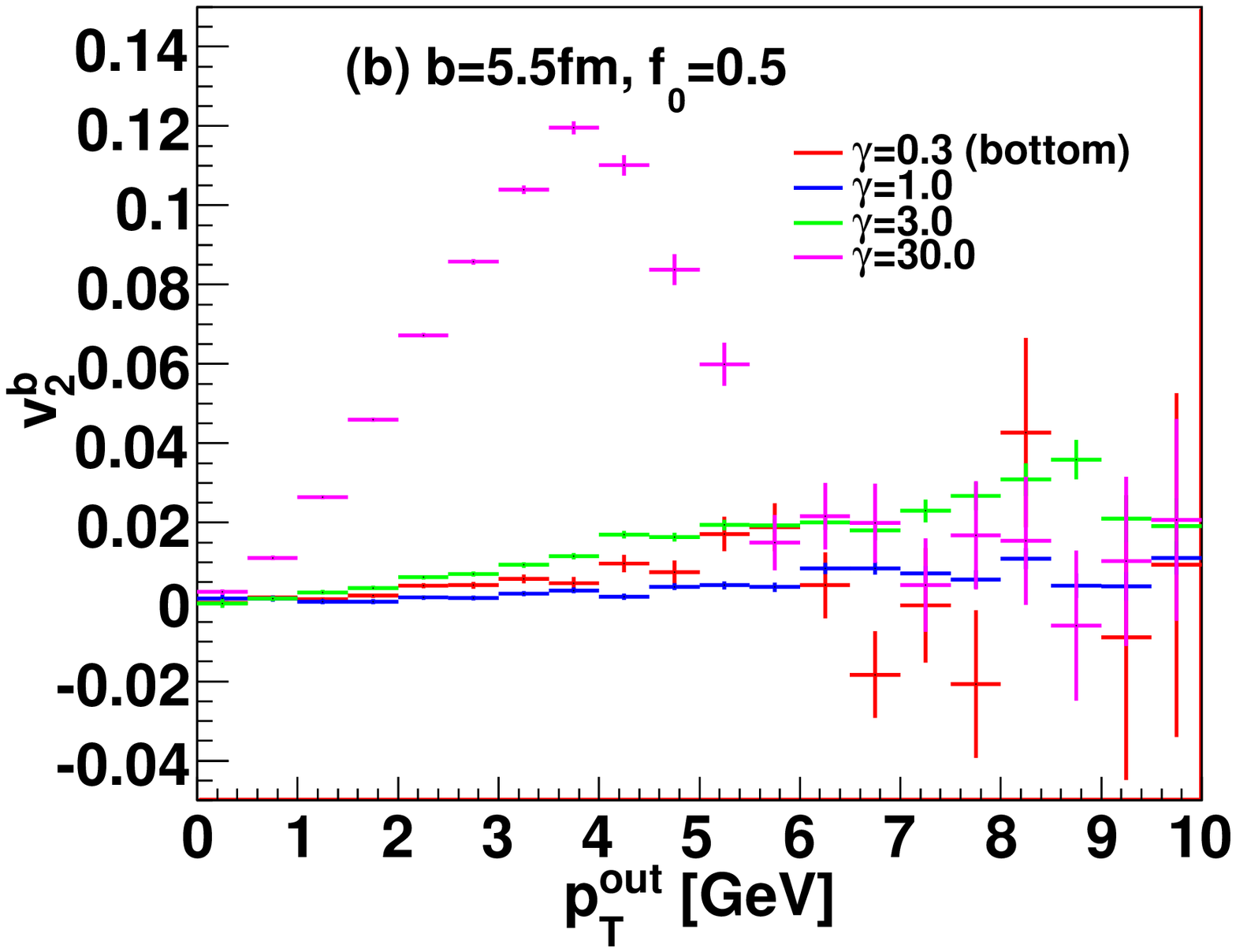}
\caption[SPECTRUM_HQ_V2]
{\footnotesize  (Color online)
 $v_{2}^Q$ of (a) charm quarks and (b) bottom quarks with drag coefficients $\gamma=$ 0.3, 1.0, 3.0, and 30.0 in mid-rapidity ($|y_{p}|\leq1.0$).
The collision geometry and the freezeout condition are the same with those in Fig.~\ref{SPECTRUM_HQ_RAA}.
In (a), the statistical errors for $\gamma=$ 3.0 and 30.0 are so large at $p_{\rm T}^{\rm out}>$ 6 GeV that we omit them.
 }
\label{SPECTRUM_HQ_V2}
\end{figure}
%\end{widetext}
In Fig.~\ref{SPECTRUM_HQ_V2}, we show the elliptic flow for the heavy quark
$v_{2}^Q$ ($Q=c, b$) at mid-rapidity at impact parameter 5.5 fm.
It is clear that the charm and bottom quarks with any drag force at 
large $p_{\rm T}^{\rm out}$ are less thermalized and thus  they do not produce much  momentum anisotropy.
Note that the dominant contributions of heavy quarks with $\gamma=$ 30.0 at large $p^{\rm out}_{\rm T}$ may be those that start near the surface of the QGP fireball and with large initial $p^{\rm in}_{\rm T}$, therefore they are not much thermalized because of the too short stay times. 
  On the other hand, charm quarks with small $p_{\rm T}^{\rm out}$ 
  are thermalized for large drag force and
  develops $v_{2}^Q$ reflecting the flow of light particles.
As for bottom quarks, they only have small momentum anisotropy with all drag forces but $\gamma=30.0$ at small $p_{\rm T}^{\rm out}$ because they are not enough thermalized.

%%%%%%%%%%%%%%%%%%%%%%%%%%%%%%%%%%%%%%%%%%%%%%%%%%%%%%%%%%%%%%%%%%%%%%%
\subsection{Electron spectra}
\label{sec4b}
%%%%%%%%%%%%%%%%%%%%%%%%%%%%%%%%%%%%%%%%%%%%%%%%%%%%%%%%%%%%%%%%%%%%%%%

%%%%%%%%%%%%%%%%%%%%%%%%%%%%%%%%%%%%%%%%%%%%%%%%%%%%%%%%%%%%%%%%%%%%%%%
\subsubsection{Nuclear modification factor $R_{\rm{AA}}$}
\label{sec4b1}
%%%%%%%%%%%%%%%%%%%%%%%%%%%%%%%%%%%%%%%%%%%%%%%%%%%%%%%%%%%%%%%%%%%%%%%
\begin{figure}
\centering
\includegraphics[width=7cm,clip]{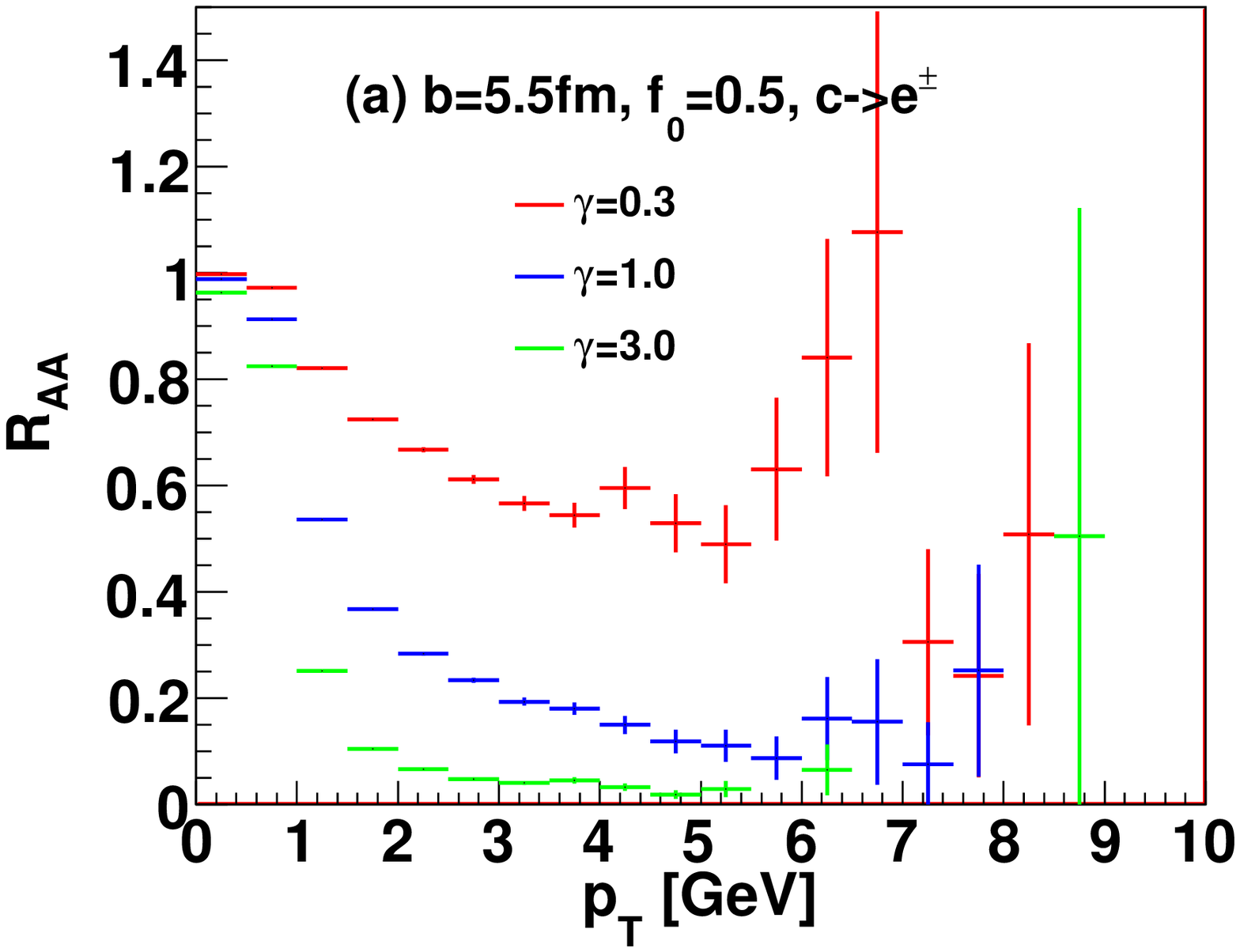}
\includegraphics[width=7cm,clip]{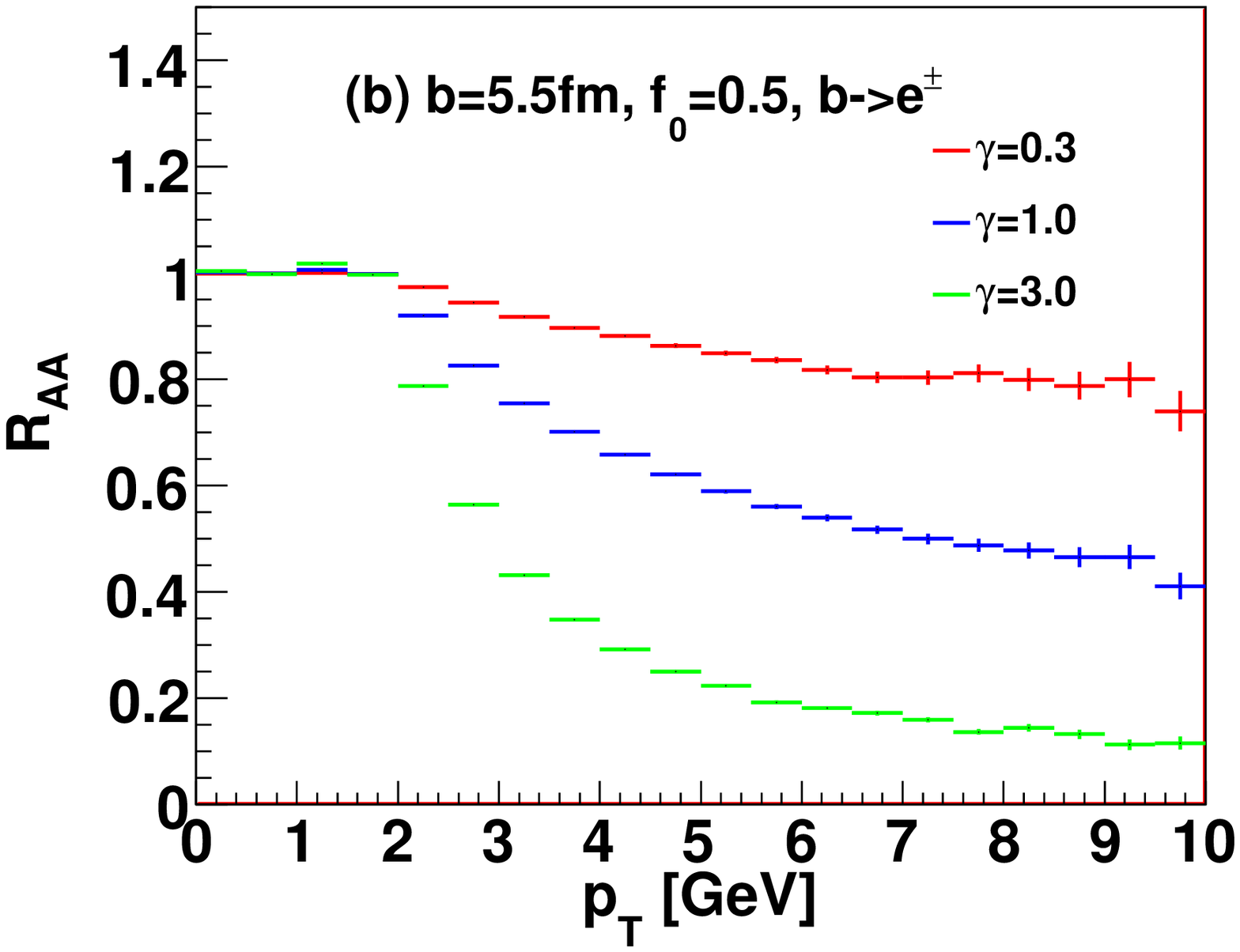}
\includegraphics[width=7cm,clip]{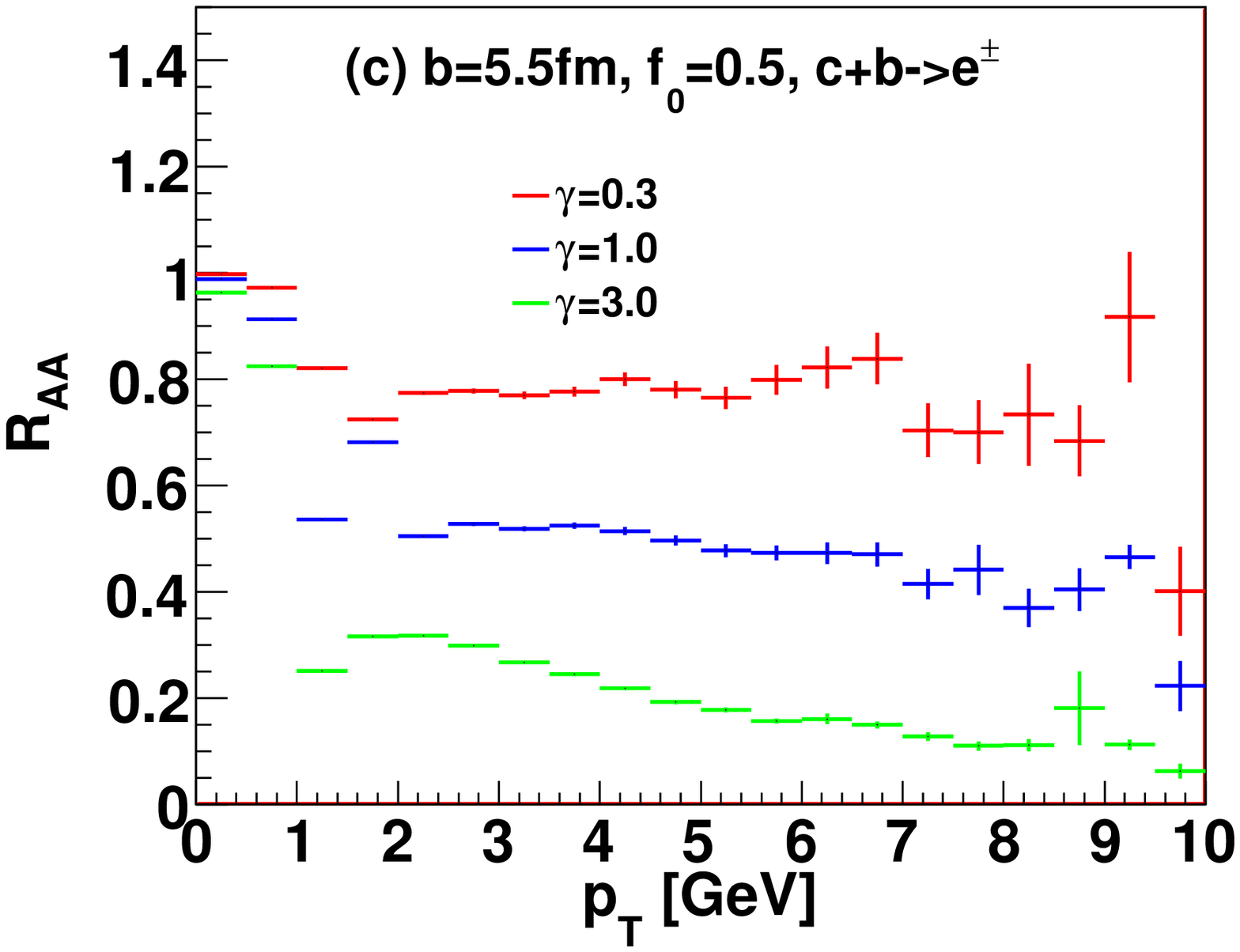}
\includegraphics[width=7cm,clip]{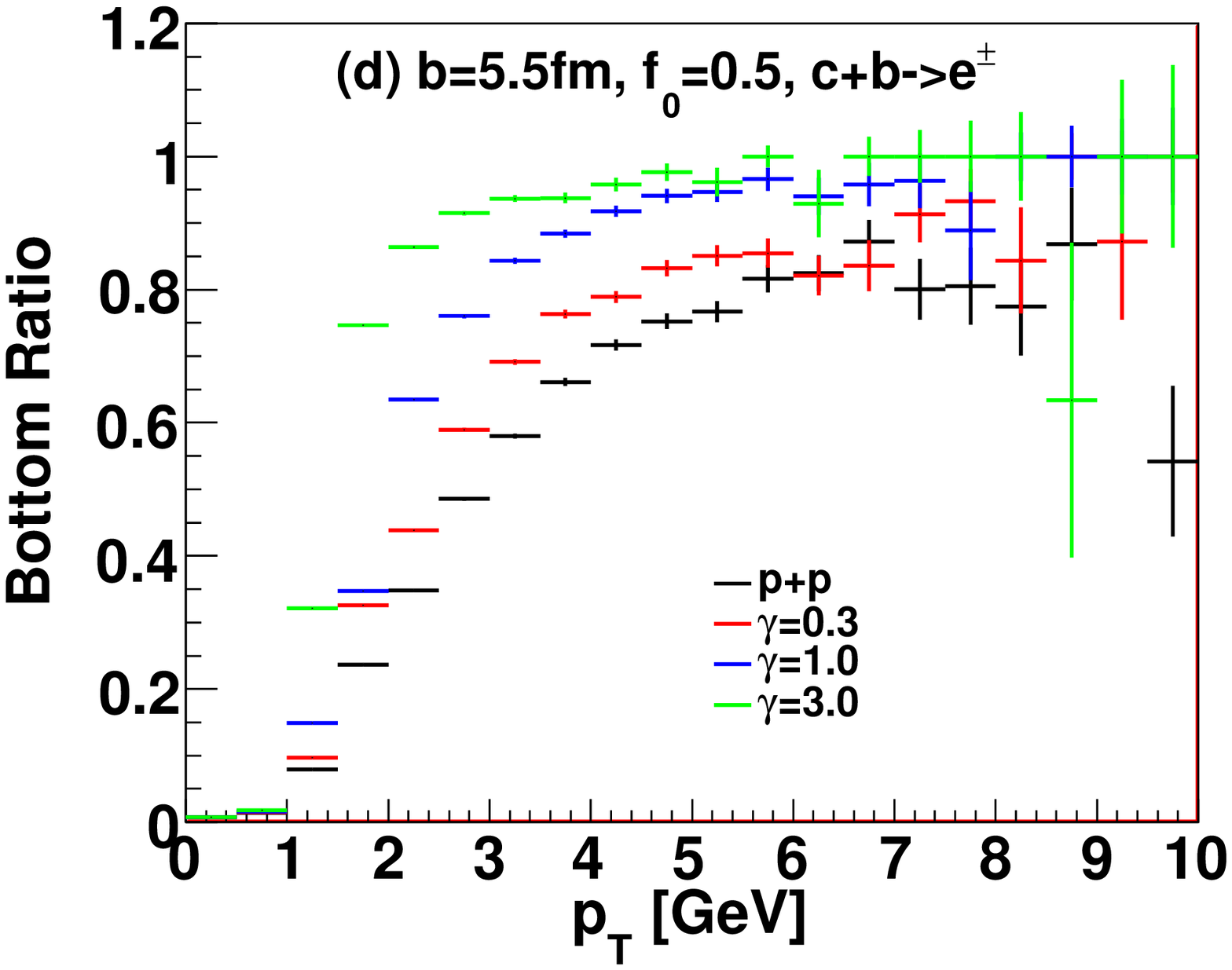}
\caption[SPECTRUM_EP_RAA]
{\footnotesize  (Color online)
(a) $R_{\rm{AA}}$ of electrons from charm,
(b) $R_{\rm{AA}}$ of electrons from bottom,
(c) $R_{\rm{AA}}$ of electrons from both charm and bottom, and
(d) the ratio of electrons from the bottom and the net electrons.
All results are in mid-rapidity ($|y_{p}|\leq0.35$).
The drag coefficient is taken to be  $\gamma=$ 0.3, 1.0, and 3.0.
 The impact parameter is taken to be 5.5 fm in Au+Au collisions. 
For freezeout condition, the $f_{0}=$ 0.5 is adopted.
In (d), the result of $p+p$ collision calculated in the leading order pQCD by PYTHIA
 is also plotted. 
}
\label{SPECTRUM_EP_RAA}
\end{figure}

 Let us now examine the results of electrons and positrons 
(we call them just electrons for short) which are the decay products from
 $D$ and/or $B$ mesons.
 In Fig.~\ref{SPECTRUM_EP_RAA}(a), (b), and (c), we 
  show $R_{\rm{AA}}$ of electrons (a) from charm quarks,
(b) from bottom quarks, and (c) from charm+bottom quarks.
 The dependence of $R_{\rm{AA}}$ on the drag coefficient $\gamma$
 is understood easily:
  Larger drag coefficient gives larger energy loss and 
   $R_{\rm{AA}}$ is suppressed.
  There is however one qualitative difference between 
  $R_{\rm{AA}}^Q$ in Sec.~\ref{sec4a2} and $R_{\rm{AA}}$
  in the low $p_{\rm T}$ region: $R_{\rm{AA}}^Q$ exceeds unity due to the 
  shift of the high momentum quarks to low momentum quarks, while
   $R_{\rm{AA}}$ stays around unity at low momentum.
  This is understood  by recognizing that the low 
  $p_{\rm T}$ electrons come from wide range of heavy quarks with various
  freezeout momenta, so that low momentum electrons
   are not sensitive to the modification of the heavy quark
  spectrum due to diffusion.
  On the other hand, the electrons with high $p_{\rm T}$ originate mainly 
   from high $p_{\rm T}$ heavy quarks and thus they are sensitive to 
   the spectral change of heavy quarks.
 
  In Fig.~\ref{SPECTRUM_EP_RAA}(d), 
  the number of electrons from bottom divided by that from charm+bottom
   for Au+Au collision is shown as a function of  electron's $p_{\rm T}$
    together with that for $p+p$ collision. In both $p+p$ and A+A,
     more than 50\% 
     of electrons come from  the bottom for $p_{\rm T} > 3 $ GeV.
      Furthermore, the ratio increases as the drag force becomes stronger.
 The kink structure of $R_{\rm{AA}}$ at $p_{\rm T}\sim$ 1 - 2 GeV in
  Fig.~\ref{SPECTRUM_EP_RAA}(c)  is understood by the fact that the 
   dominant contribution to the electrons
   changes rapidly from the charm to the bottom.

\begin{figure}
\centering
\includegraphics[width=7cm,clip]{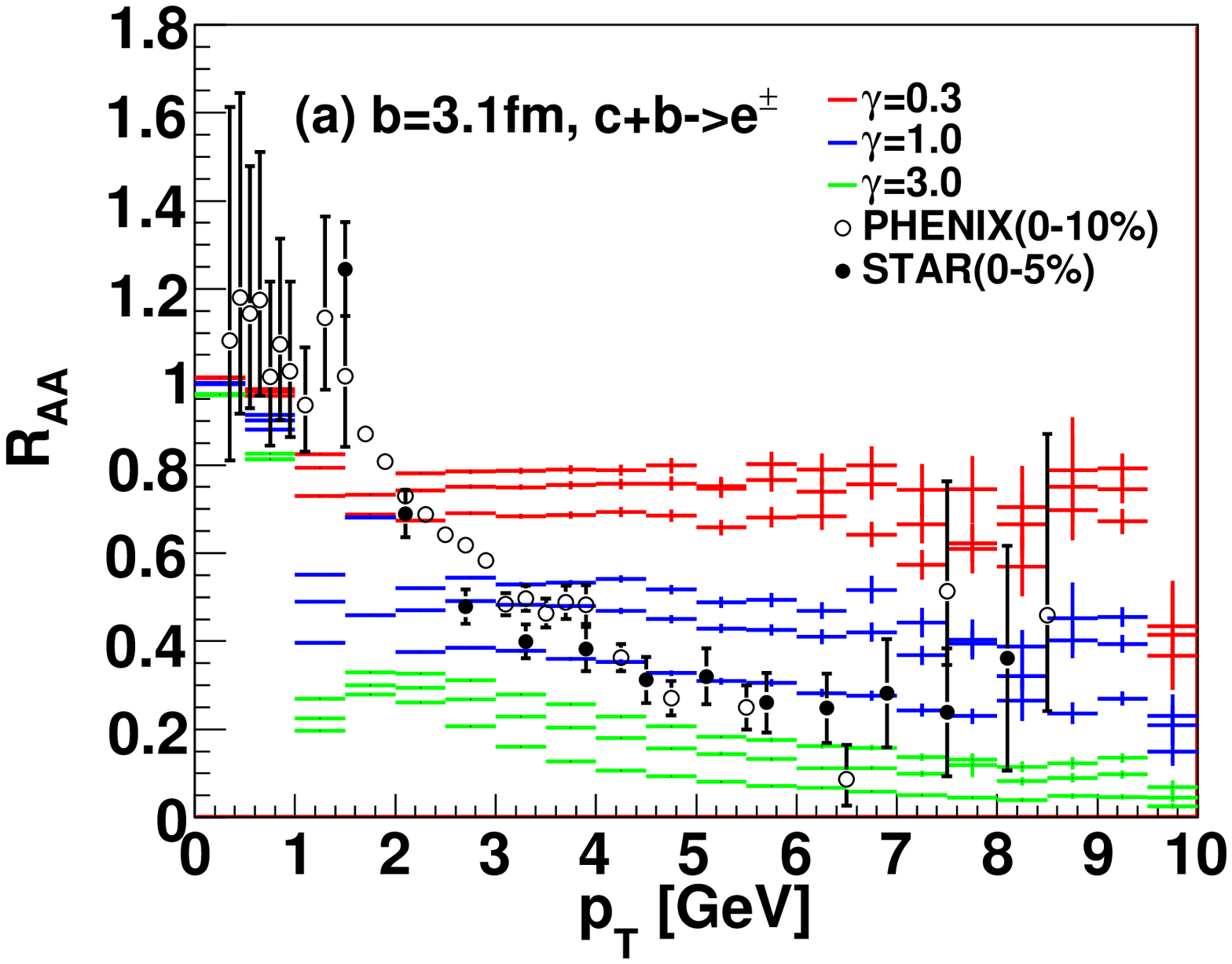}
\includegraphics[width=7cm,clip]{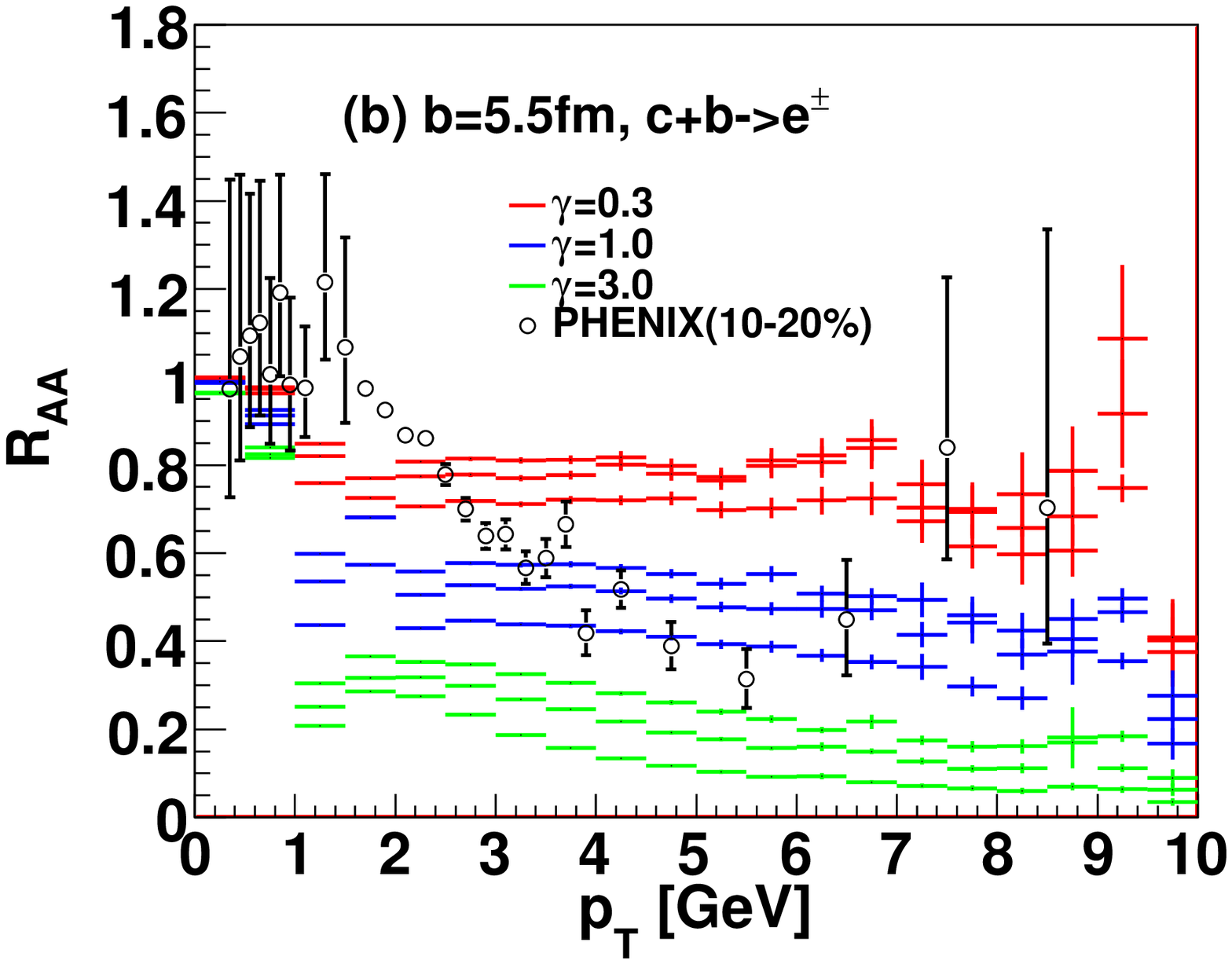}
\caption[SPECTRUM_EP_RAA_EX]
{\footnotesize  (Color online)
Comparison of  $R_{\rm{AA}}$ in our hydro + heavy-quark model
  with the experimental data \cite{phenix2007,star2007}.
 The Au+Au collision with the impact parameter (a) 3.1 fm and 
(b) 5.5 fm, both in mid-rapidity, $|y_{p}|\leq0.35$.
The drag coefficient is chosen to be $\gamma=$ 0.3, 1.0, and 3.0 
indicated by different colors.
 The freezeout condition is taken to be $f_{0}=$ 1.0, 0.5, and 0.0
 which correspond to upper, middle, and lower points, respectively,
  within the same color.
As for error bars in experimental data, we only plot the statistical errors
\cite{phenix2007,star2007}.
}
\label{SPECTRUM_EP_RAA_EX}
\end{figure}

Finally we compare our numerical results with experimental data \cite{phenix2007} in Fig.~\ref{SPECTRUM_EP_RAA_EX}.
Here we show two cases of impact parameters 3.1 fm (0-10$\%$ centrality) and 5.5 fm (10-20$\%$ centrality) at mid-rapidity.
The systematic errors due to the freezeout condition of heavy quark
 are represented by the three plots with the same color.
Recall that the comparison of our results and experimental data is only 
 reliable for $p_{\rm T} > 3$ GeV as discussed in Sec.~\ref{sec3b1} and that
  bottom quarks are the dominant source of electrons in this region.

 Although definite conclusion cannot be made from the present comparison,
 it is likely that the intermediate to large value of the  
 drag coefficient $\gamma=$ 1.0 - 3.0 is favored especially for small impact parameter. 
  This number is rather close to the value  $\gamma=2.1\pm 0.5$ predicted from
   the AdS/CFT correspondence (see Eq.~(\ref{eq:G-SYM})).
 We should remark, however, that
  the radiative energy loss \cite{Wicks2007,Armesto:2005mz}  and  the relativistic diffusion via resonances combined with quark coalescence
 \cite{vanHees:2004gq} would be legitimate alternatives to describe the data, so that
  further systematic comparison of the data and theoretical calculations is
  called for.

%%%%%%%%%%%%%%%%%%%%%%%%%%%%%%%%%%%%%%%%%%%%%%%%%%%%%%%%%%%%%%%%%%%%%%%
\subsubsection{Elliptic flow $v_{2}$}
\label{sec4b2}
%%%%%%%%%%%%%%%%%%%%%%%%%%%%%%%%%%%%%%%%%%%%%%%%%%%%%%%%%%%%%%%%%%%%%%%
\begin{figure}
\centering
\includegraphics[width=7cm,clip]{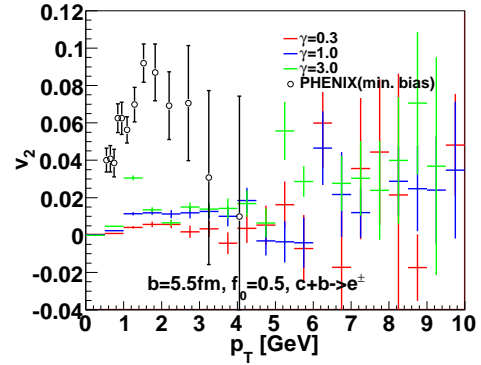}
\caption[SPECTRUM_EP_V2_EX]
{\footnotesize  (Color online)
Comparison of $v_{2}$ in our hydro + heavy-quark model
 with experimental data \cite{phenix2007} in mid-rapidity ($|y_{p}|\leq0.35$). 
Experimental data of $v_{2}$ is obtained in
 minimum bias analysis, while our theoretical values of  $v_{2}$ are
evaluated at impact parameter 5.5 fm as a representative. 
The drag coefficient is chosen to be $\gamma=$ 0.3, 1.0, and 3.0 and
the freezeout condition is $f_{0}=$ 0.5.
As for error bars in experimental data, we only plot the statistical errors
\cite{phenix2007}.
}
\label{SPECTRUM_EP_V2_EX}
\end{figure}
We show our theoretical $v_{2}$ of electrons in Fig.~\ref{SPECTRUM_EP_V2_EX}
 as a function of $p_{\rm T}$ together with 
 the experimental data \cite{phenix2007}. 
  Our $v_2$ does not depend much on the strength of the drag force
   for $p_{\rm T} > 3 $ GeV and stays small.
   Due to the poor statistic of both our simulation and the experimental data in the relevant region, it is not clear whether theory and experiment are
    consistent with each other or not. 
Although it is still preliminary, recent PHENIX data show large $v_{2}=$ 0.05 - 0.1 with small errors for $3<p_{\rm T}<5$ GeV
 at collisions with corresponding centrality \cite{phenix_new}.

%%%%%%%%%%%%%%%%%%%%%%%%%%%%%%%%%%%%%%%%%%%%%%%%%%%%%%%%%%%%%%%%%%%%%
\section{Summary and Concluding Remarks}
\label{sec5}
%%%%%%%%%%%%%%%%%%%%%%%%%%%%%%%%%%%%%%%%%%%%%%%%%%%%%%%%%%%%%%%%%%%%%%%

In this paper, we have examined the diffusion of heavy quarks in
 the dynamical QGP fluid 
 on the basis of the relativistic Langevin equation combined with the
  relativistic hydrodynamics.
 We  establish a generalized fluctuation-dissipation relation
  in It\^{o} discretization scheme, Eq.~(\ref{eq:FDT}),
 which relates the diffusion constant $D(p)$ and the drag coefficient $\Gamma(p)$
  for the relativistic Brownian particle.
  Then we parametrized the drag coefficient  motivated by the formula  
 from the AdS/CFT correspondence for strongly coupled plasma,
 $\Gamma \equiv\gamma T^{2}/M$ with the dimensionless coefficient
  $\gamma$ as a parameter.
  The space-time evolution of the QGP fluid composed of light quarks and gluons
 is treated by the  full (3+1)-dimensional  relativistic hydrodynamics
 for the perfect fluid.  
 
  By solving the Langevin equation for heavy
  quarks under the influence of the QGP fluid,
   we obtain the space-time history of the diffusion process 
  of charm and bottom in the realistic situation of the relativistic
   heavy-ion collisions.  
 The initial space-time distributions
 of charm/bottom are given by the event generator PYTHIA.
The hadronization and the semileptonic decays of charm/bottom
  after they leave from the QGP region are treated by independent
   fragmentation and decay.
  Nuclear effects for initial charm/bottom distributions
  and the quark recombination/coalescence in hadronization of heavy quarks,
  which would be important for low $p_{\rm T} < 3 $ GeV region of the final
  electron spectrum, are neglected for simplicity in this paper.
 
 Since we have the space-time history of the charm/bottom during their
 diffusion, we
 have looked at  the average stay time of heavy quarks in QGP
  $\langle t_{\rm S} \rangle $,
  the average temperature felt by heavy quarks in QGP
   $\langle \bar{T} \rangle $, and
  the average momentum loss $\langle \Delta p \rangle$ during
  the diffusion.
  We have also looked at the nuclear modification factor 
  $R_{\rm{AA}}^{Q}$ and the elliptic flow $v_{2}^Q$ of heavy quarks
    as a function of the transverse momentum
  of the heavy quarks at their freezeout $p_{\rm T}^{\rm out}$.
 The results indicate that, for sufficiently large values of 
 $p_{\rm T}^{\rm out} > 3$ GeV, there is a sizable suppression of
 $R_{\rm{AA}}^{Q}$ for large drag coefficient, while one can see
 only a significant effect in $v_2^Q$ only for  $p_{\rm T}^{\rm out} < 3$ GeV
 which is not the region one can rely on our calculation.

Then we have compared our calculations of 
$R_{\rm{AA}}$ and the elliptic flow $v_{2}$ for single electron
 with the RHIC data.  First of all, the momentum
 distribution of the electrons do not necessary reflect
  the shape of the momentum distribution of the 
  heavy quarks at their freezeout due to decay kinematics.  Also, 
 the net electrons with $p_{\rm T}> 3$ GeV
  are dominated by those from bottom quarks.
 A rough comparison of  $R_{\rm{AA}}$ for $p_{\rm T}> 3$ GeV suggests that
the drag coefficient could be as large as $\gamma=$ 1.0 - 3.0.
 On the other hand, we are unable to extract useful
  information from $v_{2}$ for $p_{\rm T}> 3$ GeV
 because of the lack of statistics in both experiment and simulations.
 The value of $\gamma=$ 1.0 - 3.0 is consistent with that
 predicted by AdS/CFT approach for strongly interacting plasma
  ($\gamma =2.1\pm 0.5$),
 although we could not exclude other descriptions of heavy quarks in QGP
  such as radiative stopping \cite{Wicks2007, Armesto:2005mz} and the resonance scattering model \cite{vanHees:2004gq}.
 High precision experimental data at RHIC and LHC for
 electrons from charm and bottom identified separately are highly 
 desirable.  Also
 the correlation of the transverse momenta  
 of a heavy quark and a heavy anti-quark (and the associated 
 electron-positron or $D$-$\bar D$ correlation \cite{Zhu et al., Zhu-Xu-Xhuang})
 could be a good observable to make detailed comparison of the
 theories and experiments.

Before closing, we remark possible improvements of our approach
 to treat the region $p_{\rm T}< 3$ GeV in a more reliable way:
  (i) Initial heavy quark distributions beyond LO need to be
   considered for better control of their absolute magnitude, 
  the $p_{\rm T}$ shape, and the charm/bottom ratio, (ii)
 nuclear effects on the initial charm/bottom distribution 
 need to be examined, and (iii) the 
 hadronization of charm/bottom due to quark recombination processes
 needs to be taken into account.

\acknowledgments
\vspace*{-2mm}
  Part of this work was carried out during 
Y. Akamatsu's visit to European Centre for Theoretical Studies in
Nuclear Physics and Related Areas ($\rm ECT^{*}$).  Y. A. thanks
 $\rm ECT^{*}$ and J.P. Blaizot for their kind hospitality.
The authors thank Tetsuo Matsui for fruitful discussions.
Y. A. is supported by JSPS fellowships for Young Scientists.
T. Hatsuda and T. Hirano are supported in part by the Grants-in-Aid of the
Japanese Ministry of Education, Culture, Sports, Science,
and Technology (Nos. 18540253, 19740130).

\appendix

\section{Relativistic Kramers equation}
In this Appendix, we derive the partial differential equation of 
$P(\vec p,\vec x,t)$ or the Kramers equation using the It\^{o} (pre-point)
 discretization scheme.
We give here the general form of the relativistic Langevin equation as
\begin{eqnarray}
\Delta \vec{x}(t)&=&\frac{p(\tilde t)}{E(p(\tilde t))}\Delta t, \nonumber \\
\Delta \vec{p}(t)&=&-A(p(\tilde t))\vec{p}(\tilde t)\Delta t \ 
+ \ \sqrt{B(p(\tilde t))}\vec{\eta}(t) \Delta t , \nonumber \\
W[\vec{\eta}(t)]d^{3}\eta(t)&=&C\cdot \exp\Bigl[-\frac{\Delta t}{2}\vec{\eta}(t)^2 \ \Bigr]d^{3}\eta(t) , \nonumber \\
\langle\eta_{i}(t)\eta_{j}(t')\rangle&=&\frac{\delta_{ij}\delta_{tt'}}{\Delta t} ,
\end{eqnarray}
where $E(p)=\sqrt{\vec p^{2}+M^{2}}$ with $M$ being the mass of the Brownian particle.
Here $\tilde t\equiv t$ corresponds to the It\^{o} discretization and
$\tilde t\equiv t+\Delta t$ corresponds to the Hanggi-Klimontovic discretization \cite{Debbasch}. 
Also $A(p)$, $B(p)$, and $\vec \eta(t)$ in the It\^{o} discretization correspond to $\Gamma(p)$, $D(p)$, and $\vec\xi(t)/(\sqrt{D(p)}\Delta t)$ in the text, respectively.
Since the Langevin equation is based on Markovian process,
one needs information only at time $t'$ in order to know the probability at later time $t$:
%\begin{widetext}
\begin{eqnarray}
&&P(\vec p,\vec x,t | \vec p_{0},\vec x_{0},t_{0}) \label{eq:Markov1} \\
&&=\int d^{3}p'd^{3}x' P(\vec p,\vec x,t | \vec p',\vec x', t')
P(\vec p',\vec x',t' | \vec p_{0},\vec x_{0},t_{0}), \nonumber
\end{eqnarray}
%\end{widetext}
where $P(X,t | X_{0},t_{0})$ ($X=\{\vec p,\vec x\}$) represents the conditional distribution function with a fixed initial 
condition $X_{0}$ at time $t_{0}$.
In order to derive the partial differential equation, we have to calculate 
$P(X,t+\Delta t | X',t)$ from the Langevin equation.
From the definition of $P(X,t+\Delta t | X',t)$,
%\begin{widetext}

\begin{eqnarray}
& &P(X,t+\Delta t | X',t) 
\equiv\langle\delta(X-X(t+\Delta t))\rangle|_{t,X'} \nonumber \\
& & \ \ \ \ \ 
=\langle\delta\bigl[X-X'-\Delta X(\eta(t),t)\bigr]\rangle  \nonumber \\
%&=&\langle\sum_{m=0}^{\infty}\bigl[-\Delta X(\eta(t),t)\bigr]^{m}\frac{1}%{m!}\partial^{m} _{X}
%\delta(X-X')\rangle 
& & \ \ \ \ \
=\sum_{m=0}^{\infty}\langle\bigl[-\Delta X(\eta(t),t)\bigr]^{m}\rangle
\frac{1}{m!}\partial^{m}_{X}\delta(X-X'), \nonumber \\
& & \langle Y(\eta(t),t) \rangle \equiv \int d^{3}\eta(t)W[\eta(t)]Y(\eta(t),t).  \label{eq:Markov2}
\end{eqnarray}
%\end{widetext}

Here $\langle\cdots \rangle|_{t,X'}$ in the first line of 
Eq.~(\ref{eq:Markov2}) represents the conditional probability with the fixed initial condition $X(t)=X'$.
Note that in the last line of Eq.~(\ref{eq:Markov2}), the average is expressed by the
variables at time $\it t$. 
Inserting Eq.~(\ref{eq:Markov2}) into Eq.~(\ref{eq:Markov1}), we obtain
%\begin{widetext}
\begin{eqnarray}
& &P(X,t+\Delta t | X_{0},t_{0}) \nonumber \\
& & \ \ 
=\int dX' P(X,t+\Delta t | X',t) P(X',t| X_{0},t_{0})\nonumber \\
& & \ \ 
=\int dX' \Bigl[ \delta(X-X')+\sum_{m=1}^{\infty}\langle\bigl[-\Delta X(\eta(t),t)\bigr]^{m}\rangle \nonumber \\
& & \ \ \ \ \ \ \ \ \ \ \ \ \ \ \
\frac{1}{m!}\partial^{m}_{X}\delta(X-X')\Bigr] 
\cdot P(X',t | X_{0},t_{0}) \nonumber \\
& & \ \
=P(X,t| X_{0},t_{0})\nonumber \\
& & \ \ \ \ \ \ \
+\sum_{m=1}^{\infty}\frac{1}{m!}\partial^{m}_{X}\Bigl[\langle\bigl[-\Delta X(\eta(t),t)\bigr]^{m}\rangle 
P(X,t | X_{0},t_{0})\Bigr] \nonumber\\
& & \ \
=P(X,t|X_{0},t_{0})+\Delta t\partial_{t}P(X,t|X_{0},t_{0}). \label{eq:formula} 
\end{eqnarray}
%\end{widetext}

In the It\^{o} discretization scheme,
the relevant average values $\langle [\Delta X(\eta(t),t)]^{m}\rangle$ are
\begin{eqnarray}
\langle \Delta \vec x(t)\rangle &=& \frac{\vec p(t)}{E(p(t))}\Delta t, \nonumber \\
\langle \Delta \vec p(t)\rangle &=& -A(p(t))\vec p(t)\Delta t, \nonumber \\
\langle \Delta p_{i}(t)\Delta p_{j}(t)\rangle
&=&B(p(t))\delta_{ij}\Delta t ,\label{eq:rel_ave}
\end{eqnarray}
and the others are in higher order in $\Delta t$.

From Eqs.~(\ref{eq:formula}) and (\ref{eq:rel_ave}), the resulting relativistic Kramers equation reads
\begin{eqnarray}
&&\Bigl(\frac{\partial}{\partial t}+\frac{\vec p}{E}\frac{\partial}
{\partial \vec x}
\Bigr)P(\vec p,\vec x,t) \nonumber\\
&&=\frac{\partial}{\partial \vec p}\Bigl(
A(p)\vec p + \frac{1}{2}\frac{\partial}{\partial \vec p}
B(p)\Bigr)P(\vec p,\vec x,t). 
\end{eqnarray}
         \\
         \\ 
         \\
         \\ 
         \\
         \\ 
         \\
         \\ 
         \\
         \\ 
         \\
         \\ 
         \\ 
         \\
         \\ 
         \\
         \\ 
         \\ 
         \\
         \\ 
         \\
         \\

%%%%%%%%%%%%%%%%%%%%%%%%  References %%%%%%%%%%%%%%%%%%%%%%%%%%%%%%%%%%%%%%%%%
\bibliographystyle{apsrev}
%\bibliography{references}

\begin{thebibliography}{99}
%QGP 
\bibitem{QGP}
K.~Yagi, T.~Hatsuda and Y.~Miake, $\it Quark$ - $\it Gluon$ $\it Plasma$
(Cambridge Univ. Press, Cambridge, 2005).
Proceedings of $\it Quark \ Matter \ 2006$,
 J.\ Phys.\ G\ {\bf 34} (2007).

%hydro review
\bibitem{Hirano:2008hy}
 T.~Hirano, N.~van der Kolk and A.~Bilandzic,
 %``Hydrodynamics and Flow,''
 arXiv:0808.2684 [nucl-th].

%hard probe
\bibitem{Vitev2008}
Reviewed in, A.~Vitev,  arXiv:0806.0003 [hep-ph].

%AuAu
\bibitem{phenix2007} 
A.~Adare {\it et al.} (PHENIX Collaboration), 
Phys.\ Rev.\ Lett.\ {\bf98}, 172301 (2007).

%AuAu
\bibitem{star2007}
  B.I.~Abelev {\it et al.}  (STAR Collaboration),
  Phys.\ Rev.\ Lett.\  {\bf 98}, 192301 (2007).


%weak coupling
\bibitem{Djordjevic2006} 
Reviewed in M.~Djordjevic, J.\ Phys.\ G {\bf 32}, S333 (2006).


%weak coupling
\bibitem{Wicks2007}
S.~Wicks, W.~Horowitz, M.~Djordjevic and M.~Gyulassy, Nucl.\ Phys.\ {\bf A784}, 426 (2007).


%weak coupling
\bibitem{Moore2008}
S.~Caron-Huot and G.D.~Moore, Phys.\ Rev.\ Lett.\ {\bf100}, 052301 (2008);
 JHEP {\bf 0802}, 081 (2008).


%ads/cft
\bibitem{Gubser2006}
S.S~.Gubser, Phys.\ Rev.\ D {\bf 74}, 126005 (2006).

%ads/cft
\bibitem{Teany2006}
J.~Casalderrey-Solana and D.~Teaney, Phys.\ Rev.\ D {\bf 74}, 085012 (2006).

%ads/cft
\bibitem{Herzog2006}
C.P.~Herzog, A.~Karch, P.~Kovtun, C.~Kozcaz and L.G.~Yaffe, 
JHEP\ {\bf 07}, 013 (2006).


%ads/cft
\bibitem{Gubser2007}
S.S.~Gubser, Phys.\ Rev.\  D {\bf 76}, 126003 (2007).

%pythia
\bibitem{Sjostrand}
  T.~Sjostrand, S.~Mrenna and P.~Skands,
  JHEP {\bf 0605}, 026 (2006).

%Langevin
\bibitem{Langevin}
 P.~Langevin, C.\ R.\ Acad.\ Sci.\ Paris\ {\bf146}, 530 (1908).\\
N.G.~Van Kampen, $\it Stochastic \ Processes \ in \ Physics$ $\it and \ Chemistry$ 
(North-Holland, Amsterdam, 1981).


%Langevin
\bibitem{Debbasch}
F.~Debbasch, K.~Mallick and J.P.~Rivet,
J. of Stat. Phys. {\bf 88}, 945 (1997).
F.~Debbasch and J.P.~Rivet, J. of Stat. Phys. {\bf 90}, 1179 (1998).\\
See also, C.~Chevalier and F.~Debbasch, 
J. Math. Phys. {\bf 49}, 043303 (2008) and references therein.

%weak coupling diffusion
\bibitem{Moore:2004tg}
  G.D.~Moore and D.~Teaney,
  Phys.\ Rev.\  C {\bf 71}, 064904 (2005).

%resonance model
\bibitem{vanHees:2004gq}
  H.~van Hees and R.~Rapp,
  Phys.\ Rev.\  C {\bf 71}, 034907 (2005). \\
See also, R.~Rapp and H.~van Hees,
  arXiv:0803.0901 [hep-ph] and references therein.

%radiation
\bibitem{Armesto:2005mz}
  N.~Armesto, M.~Cacciari, A.~Dainese, C.A.~Salgado and U.A.~Wiedemann,
  Phys.\ Lett.\  B {\bf 637}, 362 (2006).

%hydro
\bibitem{Hirano:had_cas}
T.Hirano, U.Heinz, D.Kharzeev, R.Lacey and Y.Nara, Phys.\ Let.\ B{\bf636}, 299 (2006).

%hydro
\bibitem{Hirano:1}
  T.~Hirano,
  Phys.\ Rev.\  C {\bf 65}, 011901 (2002).

%hydro
\bibitem{Hirano:2}
  T.~Hirano and K.~Tsuda,
  Phys.\ Rev.\  C {\bf 66}, 054905 (2002).

%hydro
\bibitem{Bjorken}
  J.D.~Bjorken,
  Phys.\ Rev.\  D {\bf 27}, 140 (1983).

%Hydro + something
\bibitem{HiranoCollectPapers}
T.~Hirano and Y.~Nara, Phys.\ Rev.\ C \textbf{68}, 064902 (2003); C \textbf{69}, 034908 (2004); 
Phys.\ Rev.\ Lett.\ \textbf{91}, 082301 (2003); 
M.~Isse, T.~Hirano, R.~Mizukawa, A.~Ohnishi, K.~Yoshino, and Y.~Nara, arXiv:nucl-th/0702068; 
T.~Gunji, H.~Hamagaki, T.~Hatsuda, and T.~Hirano, Phys.~Rev.~C \textbf{76}, 051901(R) (2007); 
F.-M.~Liu, T.~Hirano, K.~Werner, and Y.~Zhu, arXiv:0807.4771 [hep-ph].

%pp
\bibitem{Adare:pp}
  A.~Adare {\it et al.}  (PHENIX Collaboration),
  Phys.\ Rev.\ Lett.\  {\bf 97}, 252002 (2006).

%pp
\bibitem{Cacciari}
  M.~Cacciari, P.~Nason and R.~Vogt,
  Phys.\ Rev.\ Lett.\  {\bf 95}, 122001 (2005).

%phenix_new
\bibitem{phenix_new}
D.~Hornback, talk at $\it Quark \ Matter \ 2008$.


\bibitem{Zhu et al.}
 X.~Zhu, M.~Bleicher, S.L.~Huang, K.~Schweda, H.~Stoecker, N.~Xu and P.~Zhuang,
 Phys.\ Lett.\  B {\bf 647}, 366 (2007).

\bibitem{Zhu-Xu-Xhuang}
 X.~Zhu, N.~Xu and P.~Zhuang,
 Phys.\ Rev.\ Lett.\  {\bf 100}, 152301 (2008).

\end{thebibliography}

\end{document}